\documentclass[12pt,a4paper]{article}

\usepackage{epsfig}
\usepackage{amsmath,amsfonts,amssymb}
\usepackage{t1enc}
\usepackage{cite}

\providecommand{\openone}{\leavevmode\hbox{\small1\kern-3.8pt\normalsize1}}

\begin{document}

\vspace*{-2cm}
\begin{flushright}
UG--FT--182/05 \\
CAFPE--52/05 \\
hep-ph/0503026
\end{flushright}

\begin{center}
\begin{Large}
{\bf $\boldsymbol{\ell W \nu}$ production at CLIC: a window to \\[0.1cm]
TeV scale non-decoupled neutrinos}
\end{Large}

\vspace{0.5cm}
F. del \'Aguila \\[0.2cm] 
{\it Departamento de Física Te\'orica y del Cosmos and Centro Andaluz de Física
de Partículas elementales (CAFPE), \\
Universidad de Granada, E-18071 Granada, Spain} \\[0.4cm]
J. A. Aguilar--Saavedra \\[0.2cm] 
{\it Departamento de Física and CFTP, \\
  Instituto Superior Técnico, P-1049-001 Lisboa, Portugal} \\
\end{center}

\begin{abstract}
We discuss single heavy neutrino production $e^+ e^- \to N \nu \to \ell W \nu$,
$\ell = e,\mu,\tau$,
at a future high energy collider like CLIC, with a centre of mass energy of 3
TeV. This process could allow to detect heavy
neutrinos with masses of $1-2$ TeV if their coupling to the electron $V_{eN}$
is in the
range $0.004-0.01$. We study the dependence of the limits on the heavy neutrino
mass and emphasise the crucial role of lepton flavour in the discovery of a
positive signal at CLIC energy. We present strategies to determine heavy
neutrino properties once they are discovered, namely their Dirac or Majorana
character and the size and chirality of their charged
current couplings.  Conversely, if no signal is
found, the bound $V_{eN} \leq 0.002-0.006$ would be set for masses of $1-2$ TeV,
improving the present limit up to a factor of 30. We also extend previous work
examining in detail the flavour and mass dependence of the corresponding limits
at ILC, as well as the determination of heavy neutrino properties if they are
discovered at this collider.
\end{abstract}

\section{Introduction}

The existence of heavy neutrinos is usually associated to the see-saw mechanism
\cite{seesaw}, which provides a simple and elegant explanation for the smallness
of light neutrino masses. This economical solution has no phenomenological
implications at large colliders, however, because the new neutrinos are
extremely heavy, with masses of the order of $10^{14}$ GeV for Yukawa couplings
$Y$ of order one. These extra neutrinos also supply a mechanism to
explain the observed baryon asymmetry of the universe through leptogenesis
\cite{lepto}. 
Many attempts have been made to construct viable models of this type 
but with neutrino masses at the TeV scale \cite{tev,qd}. The price to 
pay in all cases is the loss of simplicity.
Heavy neutrinos give contributions to light neutrino
masses of the order $Y^2 v^2/m_N$, where $v$ is the vacuum expectation value of
the Higgs boson and $m_N$ the heavy neutrino mass. These contributions are far
too large for $m_N$ in the TeV range unless 
({\em i\/}) $Y$ is very small, of order $10^{-5}$, in which case the heavy
neutrino is almost
decoupled from the rest of the fermions, or ({\em ii\/}) there is another
source for neutrino masses giving a comparable contribution cancelling the
$\sim Y^2 v^2/m_N$ one from the see-saw mechanism.

Both solutions require
a theoretical effort so as to build a natural model which
reproduces light neutrino masses. In the first case, it is necessary to justify
why neutrino Yukawa couplings are much smaller than for charged leptons and
quarks. In the latter, not only it is necessary to provide an additional source
of neutrino masses, but it is also crucial to give a natural explanation for
the (apparently fine-tuned) cancellation of both contributions \cite{G}.
But despite
the disadvantage of complexity there is the significant benefit that these
models, experimentally not excluded, might be directly testable at
future colliders by searching for the
production of heavy neutrinos. (Additionally, there could be indirect evidence
of their presence in neutrino oscillation experiments \cite{BGHSZ}.)
An important question is then whether these heavy
states are indeed observable or not.
Although their masses are within the reach of forthcoming or 
planned colliders, their mixing with the Standard Model (SM) leptons 
must be also large enough to allow for their production at detectable rates. 
This is because they are SM singlets, and in the absence of new 
interactions their couplings are proportional to their mixing with the light
neutrinos.

Independently of the mass generation mechanism, heavy neutrinos
with masses of several hundreds of GeV appear in Grand Unified Theories, 
like for instance in those based on $\mathrm{SO}(10)$ or on larger groups as 
$\mathrm{E}(6)$ \cite{GUT}, and can survive to low energies \cite{BR}.
Kaluza-Klein towers of neutrinos are also predicted in models with large extra
dimensions, being possible to have the first heavy modes near the electroweak
scale \cite{AS}.
Their existence is allowed by low energy data, which
set strong constraints on their mixing with the light leptons but leave room for
their production and discovery at large colliders. If they have masses up to 400
GeV and a mixing with the electron $V_{eN} \sim 0.01$, they will be discovered at
an international linear collider (ILC) with a centre of mass (CM) energy of 500
GeV \cite{paper1}.
An eventual ILC upgrade to 800 GeV will extend the reach to higher masses,
but in order to experimentally test the existence of TeV scale neutrinos a
larger CM energy is required, which is achievable only at a future $e^+ e^-$
collider in the multi-TeV range, like the compact linear collider (CLIC) with a
CM energy of 3 TeV \cite{clic00,clic04}.

In this paper we present a study of the CLIC potential to discover heavy
neutrino singlets and determine their properties in the process $e^+ e^- \to
N \nu \to \ell W \nu$. In section \ref{sec:2} we
review the formalism and derive the interactions of heavy Dirac and Majorana
neutrino singlets with the gauge and Higgs bosons, summarising present
constraints on their couplings to the charged leptons. In section \ref{sec:3} we
discuss the general characteristics of $e W \nu$, $\mu W \nu$ and
$\tau W \nu$ final states. We analyse the different contributions to the signal
and background, stressing the crucial fact that an $eNW$ coupling is necessary
to observe the heavy neutrino in any of the channels. In section
\ref{sec:4} we describe the procedure used for our Monte Carlo calculations.
The sensitivity and limits on charged current couplings achieved at CLIC are
discussed in section \ref{sec:5}, examining also the dependence on $m_N$.
In case that a heavy neutrino was discovered, we show how its Dirac or Majorana
nature could be established and its charged couplings measured. In section
\ref{sec:6} we perform a similar analysis for neutrinos with masses of $200-400$
GeV at ILC, extending previous work \cite{paper1}, also studying the
determination of their properties and comparing with CLIC results.
In section \ref{sec:7} we draw our conclusions.

\section{Addition of neutrino singlets}
\label{sec:2}

In this paper we consider a SM extension with heavy Majorana (M) or Dirac (D)
neutrino singlets. The most common situation is that three additional heavy
eigenstates $N_i$, $i=1,2,3$ are introduced, and for definiteness this is what
we assume in this section. The formalism is however general for any number of
singlets \cite{MBLS}. In
the following we will obtain the interactions of heavy neutrinos with the light
leptons, pointing out the differences between the Dirac and Majorana cases when
they exist.

The neutrino weak isospin $T_3=1/2$ eigenstates
$\nu'_{iL}$ are the same as in the SM. In the case of Dirac neutrinos we
introduce 9 additional $\mathrm{SU}(2)_L$ singlet fields
\begin{equation}
N'_{iL} \,, \quad \nu'_{iR} \,, N'_{iR} \,, \quad \quad i=1,2,3 \,,
\quad \quad (\mathrm{D}) 
\end{equation}
which allow the light neutrinos to have Dirac masses too.
For Majorana neutrinos only three fields are added
\begin{equation}
N'_{iR} \,, \quad \quad i=1,2,3 \,, \quad \quad (\mathrm{M})
\end{equation}
with $\nu'_{iR} \equiv (\nu'_{iL})^c$, $N'_{iL} \equiv
(N'_{iR})^c$. In matrix notation, the form of the mass terms in the Lagrangian
is similar in both cases,
\begin{eqnarray}
\mathcal{L}_\mathrm{mass} & = & - \,
\left(\bar \nu'_L \; \bar N'_L \right)
\left( \! \begin{array}{cc}
\frac{v}{\sqrt 2} Y' & \frac{v}{\sqrt 2} Y \\ B' & B
\end{array} \! \right) \,
\left( \!\! \begin{array}{c} \nu'_R \\ N'_R \end{array} \!\! \right)
\; + \mathrm{H.c.} \,, \quad \quad (\mathrm{D}) \nonumber \\[0.2cm]
\mathcal{L}_\mathrm{mass} & = & - \frac{1}{2} \,
\left(\bar \nu'_L \; \bar N'_L \right)
\left( \! \begin{array}{cc}
M_L & \frac{v}{\sqrt 2} Y \\ \frac{v}{\sqrt 2} Y^T & M_R
\end{array} \! \right) \,
\left( \!\! \begin{array}{c} \nu'_R \\ N'_R \end{array} \!\! \right)
\; + \mathrm{H.c.} \,,\quad \quad (\mathrm{M})
\label{massterms}
\end{eqnarray}
where the $Y$, $B$ and $M$ blocks are $3 \times 3$ matrices.\footnote{Both
mass terms in Eqs.~(\ref{massterms}) are particular cases 
of a general $12 \times 12$ symmetric mass matrix connecting
$(\nu'_L \; (\nu'_R)^c \; N'_L \; (N'_R)^c)$ and
$(\, (\nu'_L)^c \; \nu'_R \; (N'_L)^c \; N'_R)$.
If we assign lepton numbers $L = 1$ to $\nu'_L,N'_L$ and $L = - 1$ to
$(\nu'_R)^c,(N'_R)^c$, the mass term for Dirac neutrinos corresponds to the
$L = 0$ entries (with conserved lepton number). The mass term for
Majorana neutrinos includes the Yukawa entries $Y$ with $L = 0$ and the
diagonal lepton number violating blocks $M_L$, $M_R$ with $L = 2$, $L =- 2$,
respectively, and in this case $\nu'_R$, $N'_L$ are assumed very heavy or
decoupled.}
The physical
meaning is of course different, since the $Y$ matrices correspond to Yukawa
interactions, $B$, $B'$ are bare mass terms and $M_L$, $M_R$ lepton number
violating Majorana mass matrices.\footnote{In the Dirac case the right-handed
states $\nu'_{iR}$, $N'_{iR}$ are equivalent, and by a suitable redefinition
one can always choose a weak basis with $B' = 0$. Additionally, with adequate
rotations $B$ and $M_R$ could be assumed diagonal without loss of generality.
This is not necessary for our discussion, anyway, and the results in this
section do not rely on such assumptions.}
The complete
mass matrices $\mathcal{M}$ can be diagonalised by $\mathcal{U}_L^\dagger
\mathcal{M} \, \mathcal{U}_R = \mathcal{M}_\mathrm{diag}$. For Majorana
neutrinos
$\mathcal{U}_R = \mathcal{U}_L^*$, while for Dirac neutrinos the two unitary
matrices are independent. The mass eigenstates are
\begin{equation}
\left( \!\! \begin{array}{c} \nu_L \\ N_L \end{array} \!\! \right) =
\mathcal{U}_L^\dagger
\left( \!\! \begin{array}{c} \nu'_L \\ N'_L \end{array} \!\! \right) \,,
\quad \quad 
\left( \!\! \begin{array}{c} \nu_R \\ N_R \end{array} \!\! \right) =
\mathcal{U}_R^\dagger
\left( \!\! \begin{array}{c} \nu'_R \\ N'_R \end{array} \!\! \right) \,.
\end{equation}
Both for Majorana and Dirac neutrinos the weak interaction Lagrangian is
written in the weak eigenstate basis as
\begin{eqnarray}
\mathcal{L}_W & = & - \frac{g}{\sqrt 2} \, \bar l'_L \gamma^\mu \nu'_L
W_\mu + \mathrm{H.c.} \,, \nonumber \\
\mathcal{L}_Z & = & - \frac{g}{2 c_W} \, \bar \nu'_L \gamma^\mu \nu'_L Z_\mu \,,
\end{eqnarray}
with $l'_{iL}$ the charged lepton weak eigenstates. Let us divide for
convenience the rotation matrices $\mathcal{U}_L$, $\mathcal{U}_R$ in
$3 \times 6$ blocks,
\begin{equation}
\mathcal{U}_L = \left( \! \begin{array}{c} U_L \\ U'_L \end{array} \! \right)
\,, \quad \quad
\mathcal{U}_R = \left( \! \begin{array}{c} U_R \\ U'_R \end{array} \! \right)
\,.
\end{equation}
Then, the weak interaction Lagrangian is written in the mass eigenstate basis as
\begin{eqnarray}
\mathcal{L}_W & = & - \frac{g}{\sqrt 2} \, \bar l_L \gamma^\mu \; 
U_l^\dagger U_L
\left( \!\! \begin{array}{c} \nu_L \\ N_L \end{array} \!\! \right) 
W_\mu + \mathrm{H.c.} \,, \label{ec:lw} \\
\mathcal{L}_Z & = & - \frac{g}{2 c_W} 
\left(\bar \nu_L \; \bar N_L \right)
\gamma^\mu \; U_L^\dagger U_L
\left( \!\! \begin{array}{c} \nu_L \\ N_L \end{array} \!\! \right)
Z_\mu \,,
\label{ec:lz}
\end{eqnarray}
where $U_l$ is a $3 \times 3$ unitary matrix resulting from the diagonalisation
of the charged lepton mass matrix. The extended Maki-Nakagawa-Sakata (MNS)
matrix \cite{PMNS} $V \equiv U_l^\dagger U_L$ has
dimension $3 \times 6$. Neutral interactions are described by the $6 \times 6$
matrix $X \equiv U_L^\dagger U_L$, related to the former by $X = V^\dagger V$.

The interactions with the Higgs boson $H$ are
\begin{eqnarray}
\mathcal{L}_H & = & - \frac {1}{\sqrt 2} \left( \bar \nu'_L Y N'_R
+ \bar \nu'_L Y' \nu'_R \right) \, H  + \mathrm{H.c.} \,, \nonumber \\
& = & - \frac {1}{\sqrt 2} \left(\bar \nu_L \; \bar N_L \right)
U_L^\dagger (Y U_R'+Y' \, U_R)
\left( \!\! \begin{array}{c} \nu_R \\ N_R \end{array} \!\! \right)
\, H  + \mathrm{H.c.} \,,  \quad \quad (\mathrm{D}) \nonumber \\[0.1cm]
\mathcal{L}_H & = & - \frac {1}{\sqrt 2} \; \bar \nu'_L Y N'_R \, H  
+ \mathrm{H.c.} \nonumber \\
& = & - \frac {1}{\sqrt 2}  \left(\bar \nu_L \; \bar N_L \right)
U_L^\dagger Y U_L^{'*}
\left( \!\! \begin{array}{c} \nu_R \\ N_R \end{array} \!\! \right)
 \, H  + \mathrm{H.c.} \quad \quad (\mathrm{M})
\end{eqnarray}
In order to obtain explicit expressions in terms of masses and mixing angles,
we decompose $V$ in two $3 \times 3$ blocks, $V = (V^{(\nu)} \; V^{(N)})$,
with $V^{(\nu)}$, $V^{(N)}$ parameterising the mixing of the charged leptons
with the light and heavy neutrinos, respectively. The latter is experimentally
constrained to be small (see below), thus terms of order $(V^{(N)})^2$ can be
neglected. After a little algebra, the scalar interactions of both heavy
Dirac and Majorana neutrinos can be written as
\begin{equation}
\mathcal{L}_H = - \frac {g}{2 M_W} \; \bar \nu_L V^{(\nu) \, \dagger}
\, V^{(N)} M_N N_R + \mathrm{H.c.} \,,
\end{equation}
with $M_N$ their $3 \times 3$ diagonal mass matrix $M_N = \mathrm{diag} \,
(m_{N_1},m_{N_2},m_{N_3})$. In the Dirac case there are additional Yukawa
couplings among the light neutrinos.

The mixing of heavy neutrinos with charged leptons is restricted by two groups
of processes \cite{LL,pil1,nardi,pil2,bernabeu,kagan,illana}:
({\em i\/}) $\pi \to \ell \nu$, $Z \to \nu \bar \nu$ and other tree-level
processes involving light neutrinos in the final state; ({\em ii\/})
$\mu \to e \gamma$, $Z \to \ell^+ \ell^{'-}$ and other lepton flavour violating
(LFV) processes to which heavy neutrinos can contribute at one loop
level. All these processes constrain the quantities
\begin{equation}
\Omega_{\ell \ell'} \equiv \delta_{\ell \ell'} - \sum_{i=1}^3 V_{\ell \nu_i}
V_{\ell' \nu_i}^* = \sum_{i=1}^3 V_{\ell N_i} V_{\ell' N_i}^* \,.
\label{ec:omega}
\end{equation}
The processes in the first group measure lepton charged current couplings. A
global fit yields the bounds \cite{kagan}
\begin{equation}
\Omega_{ee} \leq 0.0054 \,, \quad \Omega_{\mu \mu} \leq 0.0096 \,, \quad
\Omega_{\tau \tau} \leq 0.016 \,,
\label{eps1}
\end{equation}
with a 90\% confidence level (CL). For heavy neutrino masses in the TeV range,
LFV processes in the second group give the constraints \cite{bernabeu}
\begin{equation}
|\Omega_{e \mu}| \leq 0.0001 \,, \quad |\Omega_{e \tau}| \leq 0.01 \,, \quad
|\Omega_{\mu \tau}| \leq 0.01 \,.
\label{eps2}
\end{equation}
The limits in Eqs.~(\ref{eps1}) are model-independent to a large extent, and
independent of heavy neutrino masses as well. They imply that the mixing of the
heavy eigenstates with
the charged leptons is very small, $\sum_i |V_{\ell N_i}|^2 \leq 0.0054$,
0.0096, 0.016 for $\ell = e,\mu,\tau$, respectively. On the other hand,
the bounds in Eqs.~(\ref{eps2}) do not directly constrain the products 
$V_{\ell N_i} V_{\ell' N_i}^*$ but the sums in the r.h.s. of
Eq.~(\ref{ec:omega}), and cancellations might occur between two or more terms,
and also with other new physics contributions. These cancellations may be more
or less natural, but in any case
such possibility makes the limits in Eqs.~(\ref{eps2}) relatively
weak if more than one heavy neutrino exists \cite{paper1,GZ3}. Besides, we note that 
these limits are independent of the heavy neutrino nature. For
heavy Majorana neutrinos there is an additional restriction from the
non-observation of neutrinoless double beta decay, which is below
present experimental limits for
$|V_{eN}|^2 \leq 0.0054$ and $m_{N_i} \gtrsim 100$ GeV \cite{M}.

Since mixing of the charged leptons with heavy neutrinos is experimentally
required to be very small, the usual MNS matrix $V^{(\nu)}$ is approximately
unitary, up to corrections of
order $V_{\ell N_i}^2$. Moreover, at large collider energies the light neutrino
masses can be neglected. With these approximations $V^{(\nu)}$ can be taken
equal to the identity matrix, implying also $X_{\nu_\ell \nu_\ell'} =
\delta_{\ell \ell'}$, $X_{\nu_\ell N_i} = V_{\ell N_i}$, {\em i.e.} the vertices
between light leptons can be taken equal to their SM values for massless
neutrinos, and the couplings for flavour-changing neutral interactions
$\nu_\ell N_i Z$ are proportional to those for charged currents $\ell N_i W$.

The production of a heavy neutrino $N$ involves its interactions with the light
fermions. The charged current vertex with a charged lepton $\ell$ can
be directly read from Eq.~(\ref{ec:lw}),
\begin{equation}
\mathcal{L}_W = - \frac{g}{\sqrt 2}  \left( \bar \ell \gamma^\mu V_{\ell N}
P_L N \; W_\mu + \bar N \gamma^\mu V_{\ell N}^* P_L \ell \; W_\mu^\dagger
\right) \,.
\label{ec:lNW} 
\end{equation}
The neutral current gauge couplings with a light neutrino $\nu_\ell$ are
\begin{equation}
\mathcal{L}_Z = - \frac{g}{2 c_W}  \left( \bar \nu_\ell \gamma^\mu
V_{\ell N} P_L N + \bar N \gamma^\mu V_{\ell N}^* P_L \nu_\ell \right)
Z_\mu \,.
\label{ec:nNZ}
\end{equation}
In the Dirac case, the two terms in $\mathcal{L}_Z$ describe the interactions
of heavy neutrinos and
antineutrinos. If they are Majorana particles, the second term can be rewritten
in terms of $\bar \nu_\ell$ and $N$, giving
\begin{equation}
\mathcal{L}_Z = - \frac{g}{2 c_W} \, \bar \nu_\ell \gamma^\mu \left(  
V_{\ell N} P_L - V_{\ell N}^* P_R \right) N \; Z_\mu \,. \quad \quad
\mathrm{(M)}
\label{ec:nNZm}
\end{equation}
The scalar interactions of the heavy neutrino are
\begin{equation}
\mathcal{L}_H = - \frac{g \, m_N}{2 M_W} \, \left( \bar \nu_\ell \, V_{\ell N}
P_R N + \bar N \, V_{\ell N}^* P_L \nu_\ell \right) H \,,
\label{ec:nNH}
\end{equation}
where the second term can again be rewritten for Majorana neutrinos,
\begin{equation}
\mathcal{L}_H = - \frac{g \, m_N}{2 M_W} \, \bar \nu_\ell \left( V_{\ell N} P_R
+ V_{\ell N}^* P_L \right) N \; H \,. \quad \quad
\mathrm{(M)}
\label{ec:nNHm}
\end{equation}

For our computations it is also necessary to know the total heavy neutrino width
$\Gamma_N$. $N$ can decay in the channels $N \to W^+ \ell^-$ (if $N$ is a
Majorana fermion $N \to W^- \ell^+ $ is allowed as well), $N \to Z \nu_\ell$
and $N \to H \nu_\ell$. The partial widths for these decays are \cite{GZ3,pil3}
\begin{eqnarray}
\Gamma(N \to W^+ \ell^-) & = &  \Gamma(N \to W^- \ell^+) \nonumber \\
& = & \frac{g^2}{64 \pi} |V_{\ell N}|^2
\frac{m_N^3}{M_W^2} \left( 1- \frac{M_W^2}{m_N^2} \right) 
\left( 1 + \frac{M_W^2}{m_N^2} - 2 \frac{M_W^4}{m_N^4} \right) \,, \nonumber
\\[0.1cm]
\Gamma_D(N \to Z \nu_\ell) & = &  \frac{g^2}{128 \pi c_W^2} |V_{\ell N}|^2
\frac{m_N^3}{M_Z^2} \left( 1- \frac{M_Z^2}{m_N^2} \right) 
\left( 1 + \frac{M_Z^2}{m_N^2} - 2 \frac{M_Z^4}{m_N^4} \right) \,, \nonumber
\\[0.2cm]
\Gamma_M(N \to Z \nu_\ell) & = & 2 \, \Gamma_D(N \to Z \nu_\ell) \,, \nonumber
\\
\Gamma_D(N \to H \nu_\ell) & = &  \frac{g^2}{128 \pi} |V_{\ell N}|^2
\frac{m_N^3}{M_W^2} \left( 1- \frac{M_H^2}{m_N^2} \right)^2 \,, \nonumber
\\[0.2cm]
\Gamma_M(N \to H \nu_\ell) & = & 2 \, \Gamma_D(N \to H \nu_\ell) \,,
\label{ec:widths}
\end{eqnarray}
with $\Gamma_D$ and $\Gamma_M$ standing for the widths of a Dirac and Majorana
neutrino. The factors of two in the partial widths of $N \to Z \nu_\ell$,
$N \to H \nu_\ell$ for a Majorana neutrino are the consequence of the extra
$V_{\ell N}^*$ couplings in Eqs.~(\ref{ec:nNZm}),(\ref{ec:nNHm}), which are not
present for a Dirac neutrino.
From Eqs.~(\ref{ec:widths}) it follows that for equal values
of the mixing angles $V_{\ell N}$ the width of a heavy Majorana neutrino is
twice as large as for a Dirac neutrino. Another straightforward
consequence of these expressions is that the partial widths for $W$,
$Z$ and Higgs decays are in the ratios $2 \, : \, 1 \, : \, 1$
(the latter for $m_H \ll m_N$). Since the Higgs mass is still unknown,
we will ignore the decays $N \to H \nu_\ell$ in the calculation of $\Gamma_N$.
If these decays are included, the $W^\pm \ell^\mp$ branching ratios (and hence
the final signal cross sections) are multiplied by a factor which ranges between
$3/4$ (for $m_H \ll m_N$) and unity (for $m_H \geq m_N$).

\section{$\boldsymbol{\ell W \nu}$ production and lepton flavour}
\label{sec:3}

The existence of new heavy neutrinos is rather difficult to detect
as an excess in the total cross section for $e^+ e^- \to \ell q \bar q' \nu$.
The stringent experimental bounds on their mixing angles with the light
particles restrict the size of their contribution to this process to a few
percent, except for low $m_N$ values. Such a small increase in the cross section
is unobservable due to the inherent uncertainties in the SM prediction.
Nevertheless, the heavy neutrino contribution to this signal is dominated by
on-shell $N$ production \cite{GZ3,nprod}
$e^+ e^- \to N \nu \to \ell^- W^+ \nu \to \ell^- q \bar q' \nu$
(plus the charge conjugate) if kinematically accessible, yielding
a peak in the $\ell q \bar q'$ invariant mass distribution. If
heavy neutrino mass differences are of the order of 100 GeV or
more the neutrino peaks do not overlap, so that their experimental study can be
done independently, since in this case the interference of the relevant
amplitudes is negligible.
We will thus assume for simplicity that only one heavy neutrino $N$ is produced.
For quasidegenerate heavy neutrinos with
$(m_{N_1}-m_{N_2})/(m_{N_1}+m_{N_2}) \ll 1$ \cite{qd}
one must consider interference effects, which are not addressed here.

In $N \nu$ production the two quarks in the final state result from the decay
of an on-shell $W$ boson, hence we can safely confine the analysis to the
phase space region where their invariant mass is not far from $M_W$,
and restrict the calculation to $\ell W \nu$ production (with $W \to q \bar
q'$) in the presence of a heavy neutrino.
We first discuss the process for $\ell = e$ and later point out the
differences for $\ell = \mu,\tau$, using a reference value $m_N = 1500$ GeV.
For most purposes $\ell^-$ and $\ell^+$
production may be summed because CP-violating effects are negligible, as it is
briefly commented at the end of this section.

\subsection{Final states with electrons}
\label{sec:3.1}

\begin{figure}[htb]
\begin{center}
\begin{tabular}{ccc}
\mbox{\epsfig{file=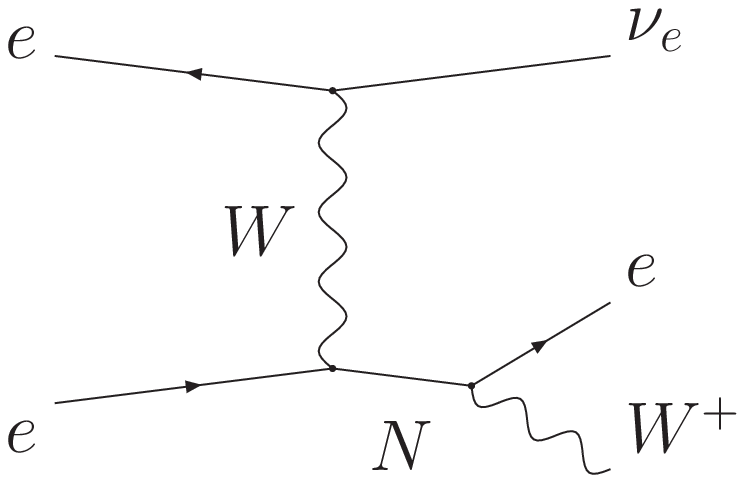,height=2.5cm,clip=}} &
\mbox{\epsfig{file=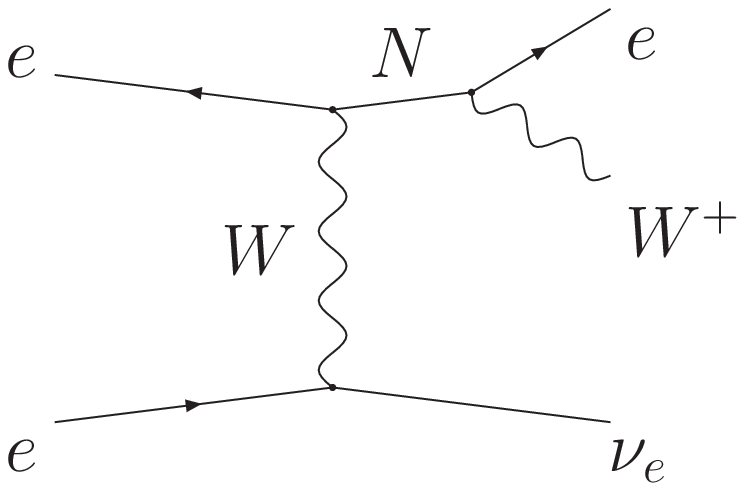,height=2.5cm,clip=}} &
\mbox{\epsfig{file=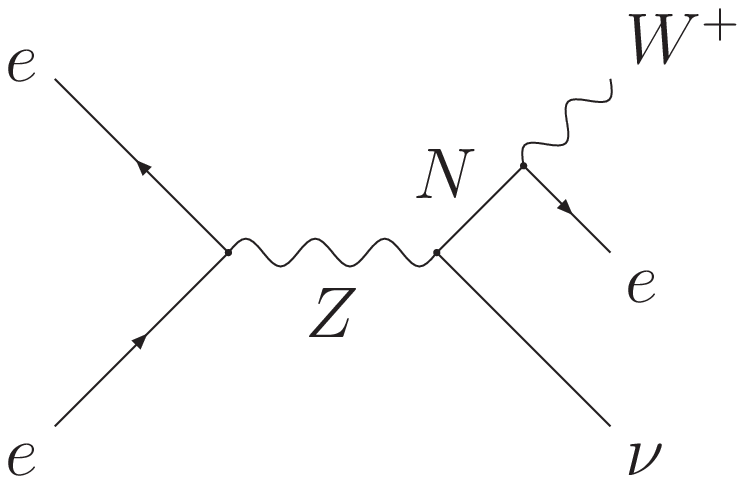,height=2.5cm,clip=}} \\
\ref{fig:diagN}a & \ref{fig:diagN}b & \ref{fig:diagN}c \\
\mbox{\epsfig{file=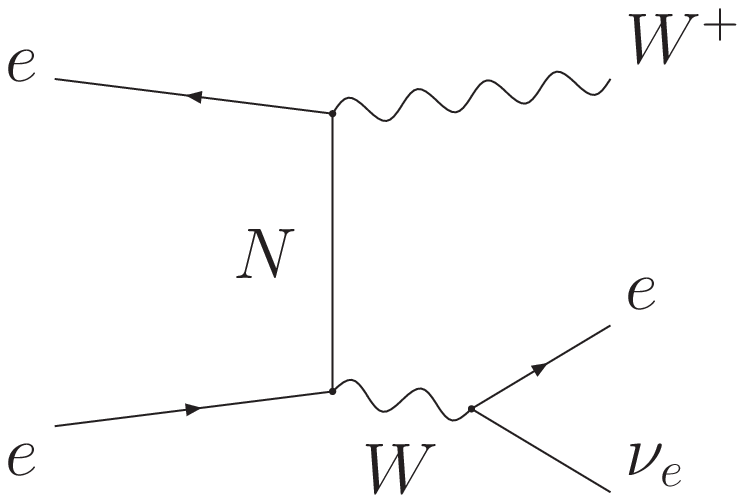,height=2.5cm,clip=}} &
\mbox{\epsfig{file=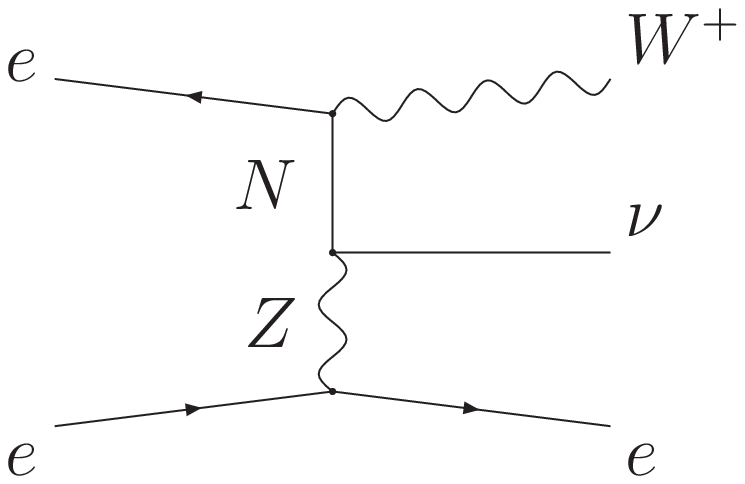,height=2.5cm,clip=}} &
\mbox{\epsfig{file=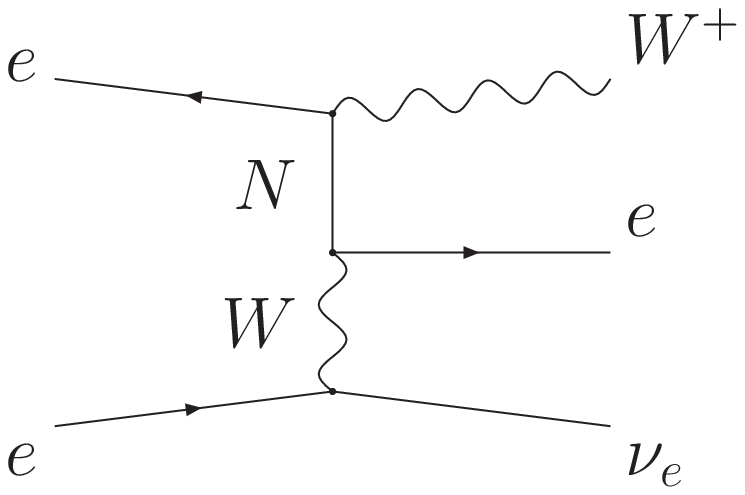,height=2.5cm,clip=}} \\
\ref{fig:diagN}d & \ref{fig:diagN}e & \ref{fig:diagN}f
\end{tabular}
\end{center}
\caption{Feynman diagrams for $e^+ e^- \to e^- W^+ \nu$ mediated by a heavy
neutrino $N$. The differences between the Dirac and Majorana cases are
discussed in the text.}
\label{fig:diagN}
\end{figure}

The Feynman diagrams for $e^+ e^- \to e^- W^+ \nu$ involving
heavy neutrino exchange are shown in Fig.~\ref{fig:diagN}. We neglect the
electron mass (as well as light quark masses) in the computation of matrix
elements, thus we do not include scalar diagrams. When the outgoing
light neutrino flavour is constrained to be $\nu_e$ we write it explicitly.
Diagrams \ref{fig:diagN}b and \ref{fig:diagN}f are present only if $N$ is
a Majorana fermion, while the rest are common to the Dirac and Majorana cases.

We notice that the six diagrams have an $eNW$ vertex, whose presence is thus
necessary for this process to occur.
The three first ones involve the production of on-shell $N$, the
corresponding amplitudes being proportional to $V_{eN}$ times a factor
$O(1)$ from the $N$ decay branching ratio. Their contribution to the cross
section is then quadratic in $V_{eN}$ (interference with the SM
amplitude is negligible). The last three diagrams
are proportional to $V_{eN}^2$, giving $V_{eN}^2$ terms in the
cross
section (through interference), plus $V_{eN}^4$ terms.
With present limits on $V_{eN}$, the former are one order of magnitude larger
than the latter, but still remain two orders of magnitude below the size of
$V_{eN}^2$ terms from diagrams with $N$ on its mass shell.
Among these, the $t$- and $u$-channel $W$ exchange diagrams (\ref{fig:diagN}a
and \ref{fig:diagN}b, respectively) are the only ones relevant at CLIC energy
and $s$-channel $Z$ exchange (\ref{fig:diagN}c) is highly suppressed, being
a factor $\sim 5$ smaller than off-shell contributions.
We then arrive to one crucial point: for equal values of the mixing
angles, the contributions of Majorana and Dirac heavy neutrinos to the
$e^- q \bar q' \nu$ cross section are almost equal.
The reason is simple: for a Majorana $N$ the neutrino signal is strongly
dominated by two non-interfering Feynman diagrams which give equal contributions
to the cross section, instead of one in the Dirac case. On the other hand, the
width of a Majorana $N$ is twice as large as for a Dirac one, as noted in the
previous section.
 
\begin{figure}[htb]
\begin{center}
\begin{tabular}{ccc}
\mbox{\epsfig{file=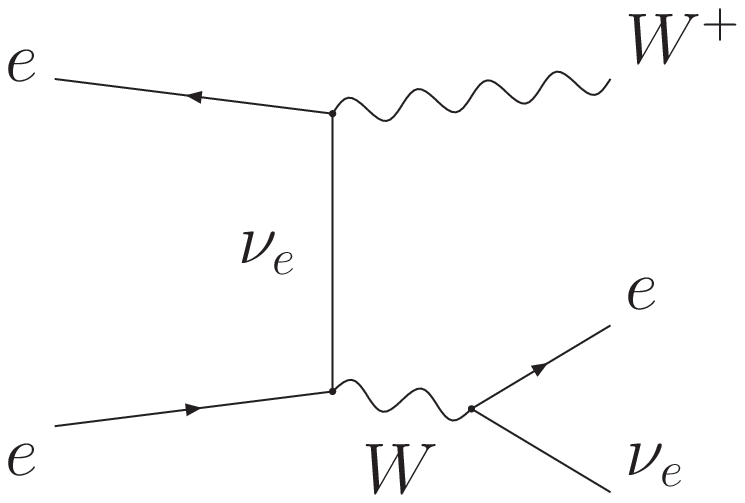,height=2.5cm,clip=}} &
\mbox{\epsfig{file=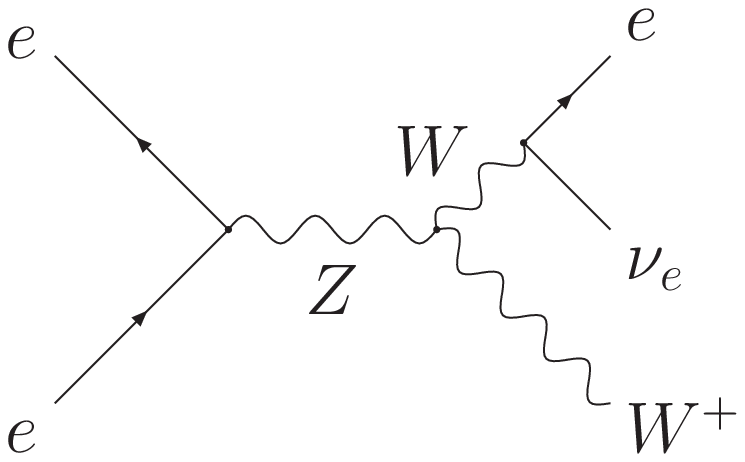,height=2.5cm,clip=}} &
\mbox{\epsfig{file=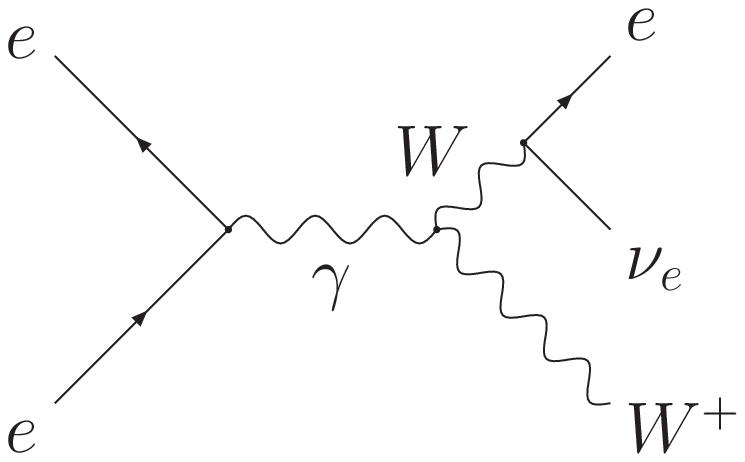,height=2.5cm,clip=}} \\
\ref{fig:diagSM}a & \ref{fig:diagSM}b & \ref{fig:diagSM}c \\
\mbox{\epsfig{file=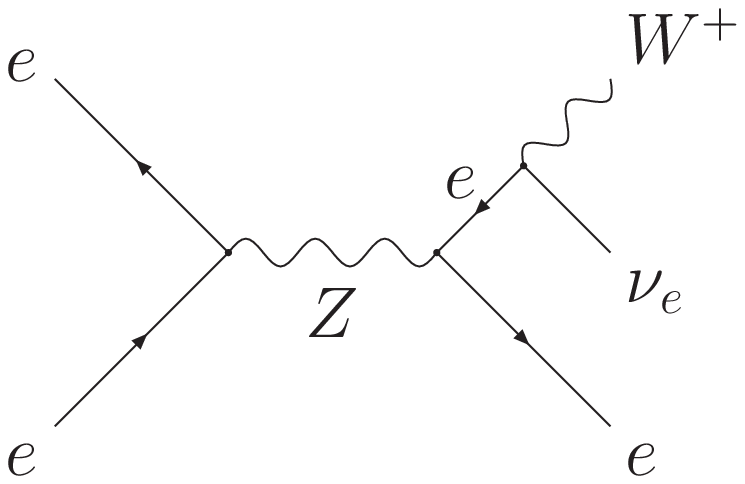,height=2.5cm,clip=}} &
\mbox{\epsfig{file=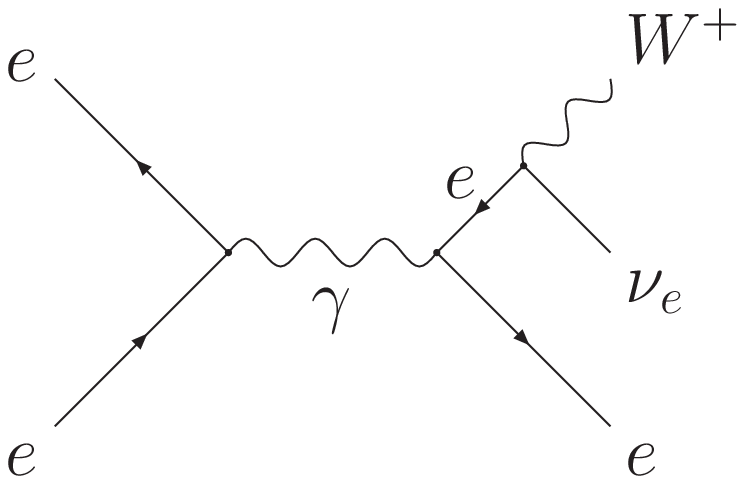,height=2.5cm,clip=}} &
\mbox{\epsfig{file=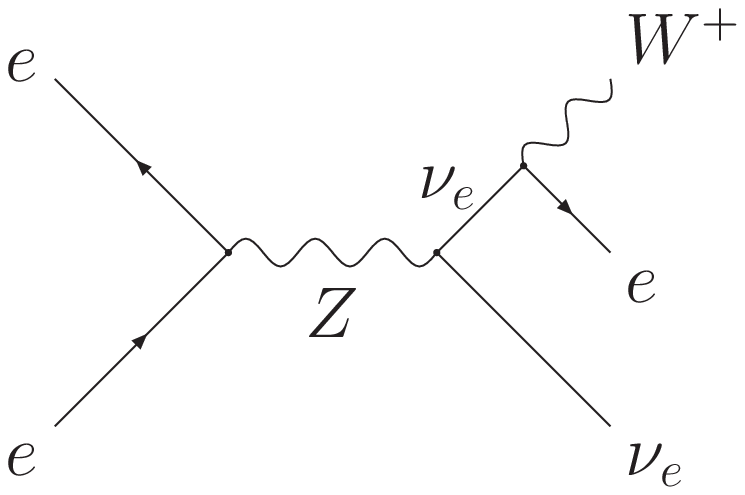,height=2.5cm,clip=}} \\
\ref{fig:diagSM}d & \ref{fig:diagSM}e & \ref{fig:diagSM}f \\
\mbox{\epsfig{file=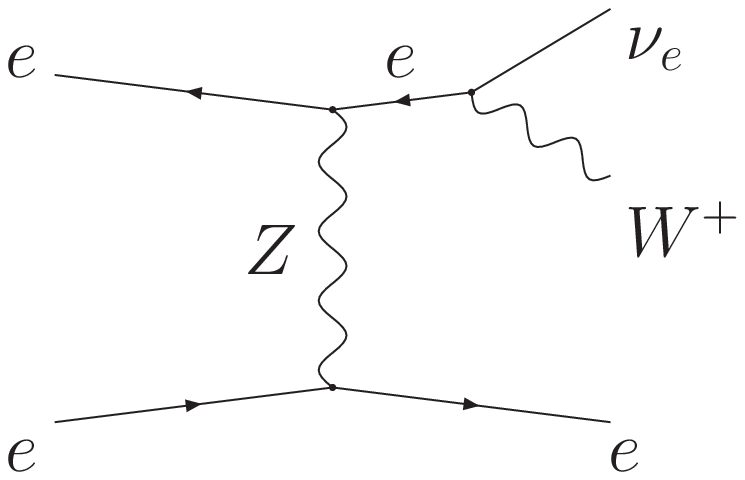,height=2.5cm,clip=}} &
\mbox{\epsfig{file=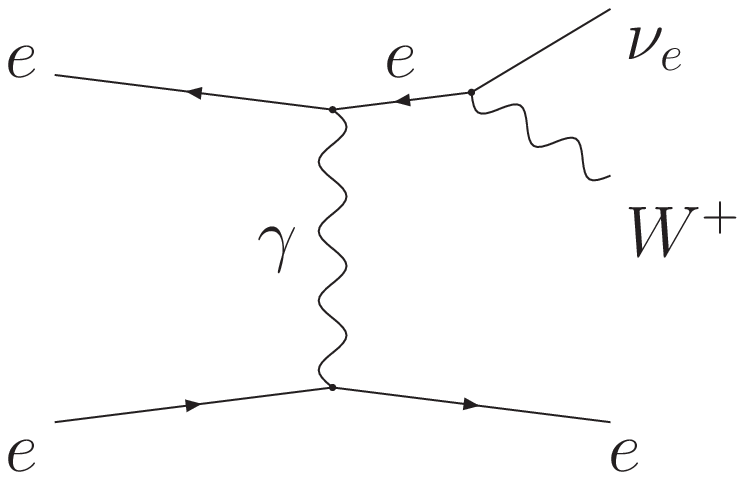,height=2.5cm,clip=}} &
\mbox{\epsfig{file=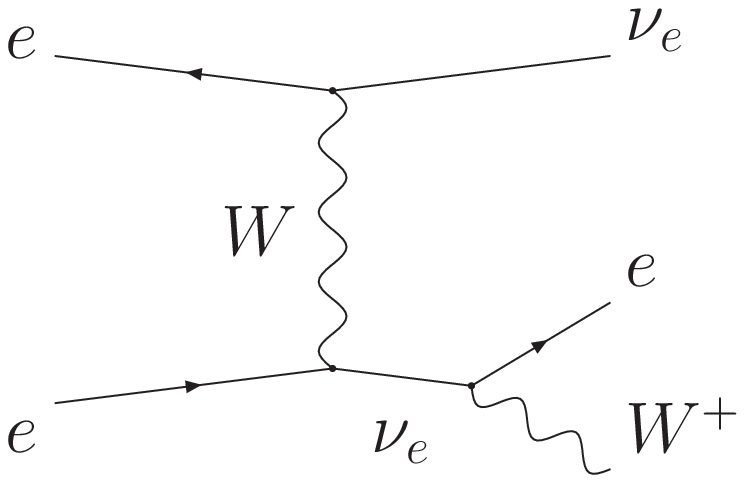,height=2.5cm,clip=}} \\
\ref{fig:diagSM}g & \ref{fig:diagSM}h & \ref{fig:diagSM}i \\
\mbox{\epsfig{file=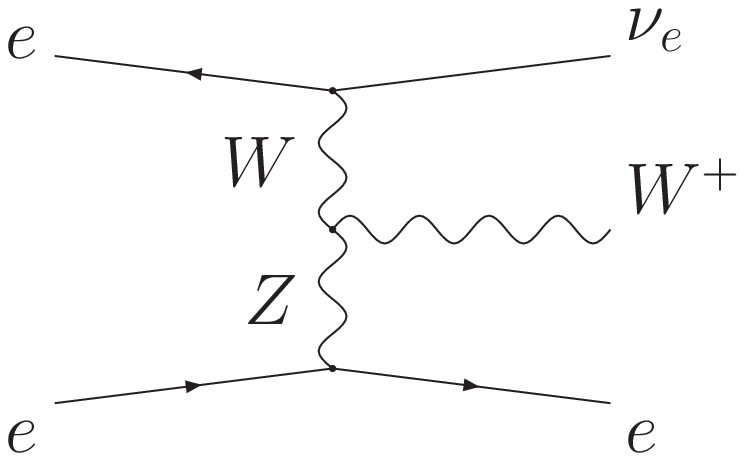,height=2.5cm,clip=}} &
\mbox{\epsfig{file=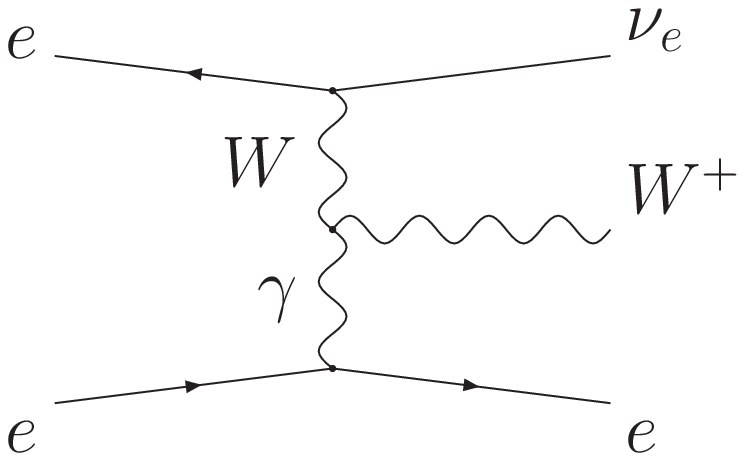,height=2.5cm,clip=}} &
\mbox{\epsfig{file=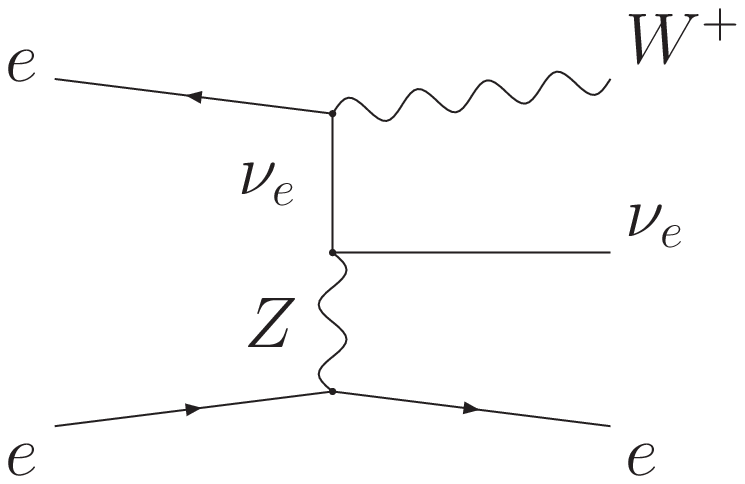,height=2.5cm,clip=}} \\
\ref{fig:diagSM}j & \ref{fig:diagSM}k & \ref{fig:diagSM}l
\end{tabular}
\end{center}
\caption{Feynman diagrams for $e^+ e^- \to e^- W^+ \nu_e$ within the SM.}
\label{fig:diagSM}
\end{figure}

In the phase space region of interest, the relevant SM diagrams are those for
$e^+ e^- \to e^- W^+ \nu_e$ with subsequent hadronic $W$ decay, as depicted in
Fig. \ref{fig:diagSM}.  The main contribution comes from diagram
\ref{fig:diagSM}i and is one order of magnitude larger than the rest. This fact
constrasts with the behaviour at ILC energy, where about one half of the cross
sections comes from resonant $W^+ W^-$ production, especially from diagram
\ref{fig:diagSM}a. The 8 additional diagrams for
$e^+ e^- \to e^- q \bar q' \nu$ which do not involve the decay
$W^+ \to q \bar q'$ give an irrelevant contribution (smaller than $0.5$ \%)
in the phase space region studied. The quadratic corrections to the
$\ell \nu W$ and $\nu \nu Z$ vertices neglected in section \ref{sec:2}
are unobservable with the available statistics. We also ignore
diagrams like \ref{fig:diagN}b and \ref{fig:diagN}f with the exchange of a
light Majorana
neutrino, which are suppressed by the ratio $m_\nu/Q \sim 10^{-13}$, where
$Q \sim \sqrt s$ is the typical energy scale in the process.

It is worth pointing out the effect of beam polarisation on the cross sections
for $e^- W^+ \nu$ production through $N$ exchange only and
through SM diagrams. For $N$ exchange we have
$\sigma_{e_R^+ e_L^-} \,:\, \sigma_{e_L^+ e_R^-} = 44000 \,:\, 1$ (for $m_N =
1500$ GeV, $V_{eN} = 0.05$), with
$\sigma_{e_R^+ e_R^-} = \sigma_{e_L^+ e_L^-} = 0$. For the SM process,
we find $\sigma_{e_R^+ e_L^-} \,:\, \sigma_{e_R^+ e_R^-} \,:\,
\sigma_{e_L^+ e_R^-} = 3700 \,:\, 200 \,:\, 1$, $\sigma_{e_L^+ e_L^-} = 0$.
Therefore, the use of negative electron polarisation $P_{e^-}$ and positive
positron polarisation $P_{e^+}$ improves the observability of the signal as well
as the statistics. Besides, it can be seen that $e_L^+ e_R^-$ cross sections
only receive contributions from diagrams with $Z$ or photon $s$-channel
exchange, thus the huge difference between $e_L^+ e_R^-$ and $e_R^+ e_L^-$ cross
sections reflects the suppression of the former.

\subsection{Final states with muons and taus}
\label{sec:3.2}

\begin{figure}[htb]
\begin{center}
\begin{tabular}{ccc}
\mbox{\epsfig{file=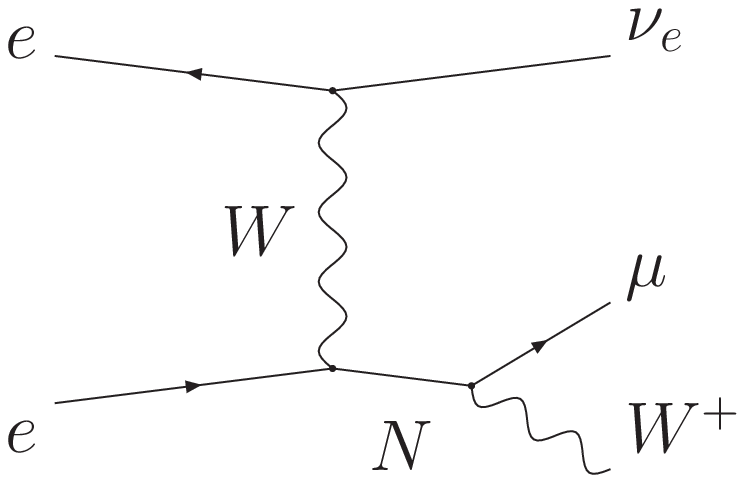,height=2.5cm,clip=}} &
\mbox{\epsfig{file=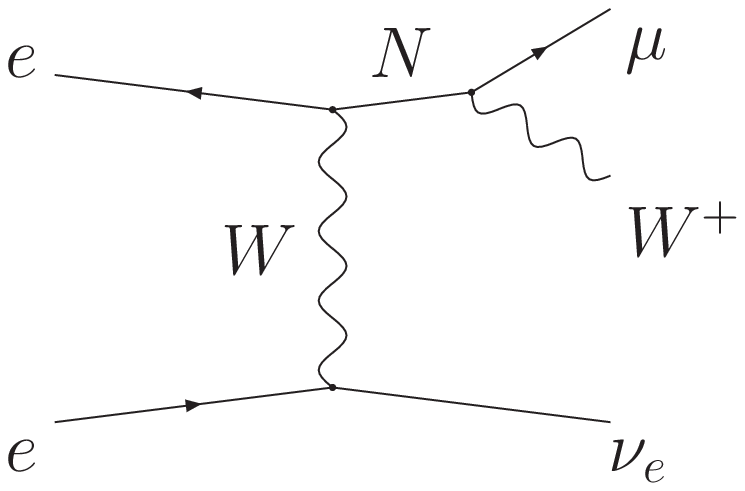,height=2.5cm,clip=}} &
\mbox{\epsfig{file=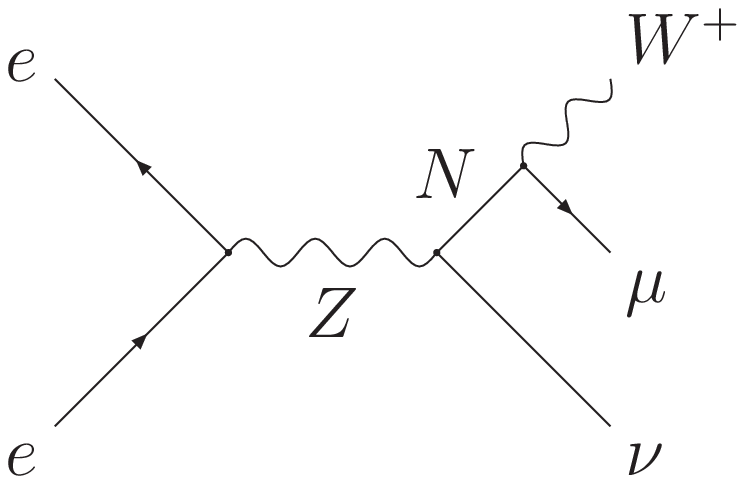,height=2.5cm,clip=}} \\
\ref{fig:diagN2}a & \ref{fig:diagN2}b & \ref{fig:diagN2}c \\
\mbox{\epsfig{file=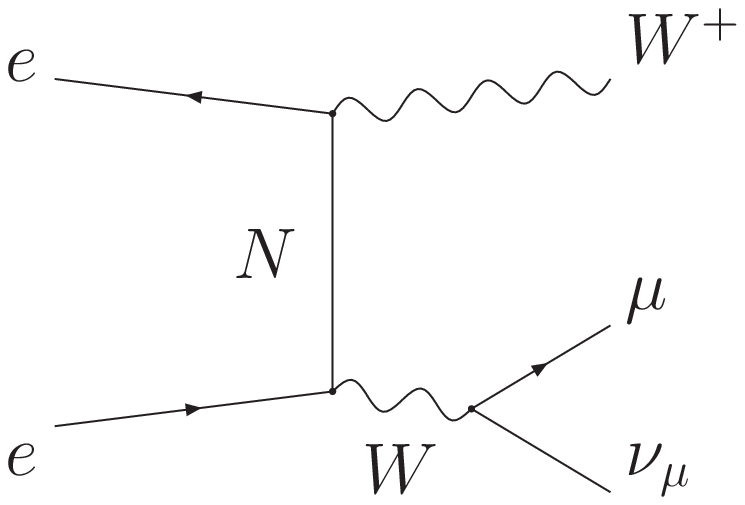,height=2.5cm,clip=}} &
\mbox{\epsfig{file=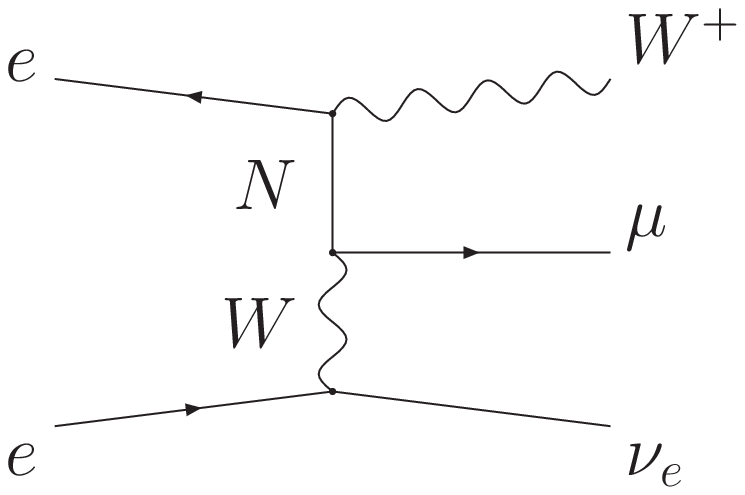,height=2.5cm,clip=}} \\
\ref{fig:diagN2}d & \ref{fig:diagN2}e
\end{tabular}
\end{center}
\caption{Feynman diagrams for $e^+ e^- \to \mu^- W^+ \nu$ mediated by a heavy
neutrino $N$. The differences between the Dirac and Majorana cases 
are discussed in the text.}
\label{fig:diagN2}
\end{figure}

Final states with $\ell=\mu,\tau$ share similar production properties and we
refer to muons for brevity. The diagrams for $e^+ e^- \to \mu^- W^+ \nu$
via heavy neutrino exchange are shown in Fig.~\ref{fig:diagN2}. The same
comments regarding the contributions of the diagrams in Fig.~\ref{fig:diagN}
apply in this case (up to different mixing angles). We observe that all
contributions except \ref{fig:diagN2}c involve an electron-heavy neutrino
interaction; in
particular, the leading diagrams \ref{fig:diagN2}a and \ref{fig:diagN2}b
correspond to heavy neutrino production via an $eNW$ vertex with subsequent
decay through a $\mu NW$ interaction. This
leads to the important consequence that the $\mu^- W^+ \nu$ signal of
heavy neutrinos is relevant only if $N$ simultaneously mixes with the electron
and muon.
We also notice that, even without mixing with the muon, a heavy neutrino can
mediate  $\mu^- W^+ \nu$ production, via diagram \ref{fig:diagN2}d.
Nevertheless, the cross section is very small in this case.
The SM background is $e^+ e^- \to \mu^- W^+ \nu_\mu$, with 6 Feynman
diagrams like the ones in \ref{fig:diagSM}a--\ref{fig:diagSM}f but replacing
$e^-$, $\nu_e$ by $\mu^-$, $\nu_\mu$, respectively. Its cross section is
dominated by resonant $W^+ W^-$ production and is 30 times smaller than for
$e^- W^+ \nu_e$.

In order to discuss the effect of beam polarisation we have to distinguish two
cases. If $\mu^- W^+ \nu$ production takes place mainly through a $eNW$
coupling, beam
polarisation has clearly the same effect as in $e^- W^+ \nu$ production:
negative $P_{e^-}$ and positive $P_{e^+}$ enhance the signal and background
and thus improve the statistics. If the heavy neutrino does not mix
with the electron but mixes with the muon, the
reverse situation occurs. Since the only contribution comes from diagram
\ref{fig:diagN2}c, the use of left-handed positrons and
right-handed electrons actually increases the signal, while reducing the SM
cross section for this process \cite{paper1}. This case has no interest at CLIC
energy, anyway, because for $V_{eN}=0$ the signal is not observable.

\subsection{CP violation with heavy neutrinos}

In this work we do not address any CP violation effects, which are unobservable
in the processes studied.
Partial rate asymmetries between $\ell^-$ and $\ell^+$ final states are
negligible at the tree level. They require interference of diagrams with
different CP-conserving phases ({\em e.g.} diagram \ref{fig:diagN}a, with a
phase arising from the $N$ propagator \cite{eilam}, and one of the diagrams in
Fig.~\ref{fig:diagSM}).
This interference is very small due to kinematics (the width $\Gamma_N$ is small
compared to the energy scale). Another possibility is the
study of triple-product asymmetries in the decay of the heavy neutrino. However,
these asymmetries are proportional to the mass of the final state charged
lepton \cite{sneutrino}, and hence very small. Therefore, we sum $\ell^-$ and
$\ell^+$ final states in all our results, unless otherwise stated.

\section{Event generation}
\label{sec:4}

The matrix elements for $e^+ e^- \to \ell^- W^+ \nu \to \ell^- q \bar q' \nu$
are calculated using {\tt HELAS} \cite{helas}, including all spin correlations
and finite width effects. We sum SM and heavy neutrino-mediated diagrams at
the amplitude level. For Majorana fermions the Feynman rules are given
in Ref.~\cite{denner}. We assume a CM energy of 3 TeV, with a beam spread of
0.35\% \cite{clic00}, ignoring an eventual beam crossing angle of 0.02 rad
at the interaction point \cite{clic04} whose effect in our simulation is
very small. We make use of electron polarisation $P_{e^-} = -0.8$
and positron polarisation $P_{e^+} = 0.6$.
The luminosity is taken as 1000 fb$^{-1}$ per year. In order to take into
account the effect of initial state radiation (ISR) and beamstrahlung we
convolute the differential cross section with ``structure functions''
$D_\mathrm{ISR}(x)$, $D_\mathrm{BS}(x)$,
\begin{equation}
d \sigma =  \int_0^1 d\sigma(x_1 y_1 E,x_2 y_2 E) D_\mathrm{ISR}(x_1)
D_\mathrm{BS}(y_1) D_\mathrm{ISR}(x_2) D_\mathrm{BS}(y_2) \,
dx_1 dy_1 dx_2 dy_2 \,.
\end{equation}
The function describing the effect of ISR used is \cite{isr}
\begin{eqnarray}
D_\mathrm{ISR}(x) & = & \frac{\eta}{2} (1-x)^{\frac{\eta}{2}-1}
\frac{e^{\frac{\eta}{2} \left( \frac{3}{4}-\gamma
\right)}}{\Gamma \left( 1 + \frac{\eta}{2} \right)}
\nonumber \\
& & \times \left[ \frac{1}{2} ( 1+x^2 ) - \frac{\eta}{8}
\left( \frac{1}{2} (1+3x^2) \log x - (1-x)^2 \right) \right] \,,
\end{eqnarray}
where
\begin{equation}
\eta(s) = -6 \log \left[ 1- \frac{\alpha_0}{3 \pi} \log \frac{s}{m_e^2} \right]
\,,
\end{equation}
$\gamma$ is the Euler constant, $\alpha_0=1/137$ the fine structure
constant, $s$ the center of mass energy squared and $m_e$ the electron mass.
For beamstrahlung we use \cite{peskin}
\begin{equation}
D_\mathrm{BS}(x) = e^{-N} \left[ \delta(x-1) +
\frac{e^{-\kappa(1-x)/x}}{x(1-x)} h(y) \right] \,,
\end{equation}
with $N=N_\gamma /2$, $\kappa=2/3 \, \Upsilon$. For the proposed luminosity
we take the parameters $\Upsilon = 8.1$, $N_\gamma = 2.3$
\cite{clic00}. The function $h(y)$ is \cite{BS2}
\begin{equation}
h(y) = \sum_{n=1}^{\infty} \frac{y^n}{n! \, \Gamma(n/3)} \,,
\end{equation}
where $y=N[\kappa \, (1-x)/x]^{1/3}$. For large $y$, $h(y)$ has the asymptotic
expansion
\begin{equation}
h(y) = \left( \frac{3z}{8\pi} \right)^\frac{1}{2} e^{4z}
\left[ 1- \frac{35}{288 \, z} - \frac{1295}{16588 \, z^2} + \cdots \right] \,,
\end{equation}
with $z=(y/3)^{3/4}$.

In final states with $\tau$ leptons, we select $\tau$ decays to $\pi$, $\rho$
and $a_1$ mesons (with a combined branching fraction of 54\% \cite{PDB}), in
which a single
$\nu_\tau$ is produced, discarding other hadronic and leptonic decays.
We simulate the $\tau$ decay assuming that the meson and $\tau$ momenta are
collinear (what is a good approximation for high $\tau$ energies) and assigning
a random fraction $x$ of the $\tau$ momentum to the meson, according to the
left-handed probability distributions \cite{taudecays}
\begin{equation}
P(x) = 2 (1-x)
\end{equation}
for pions, and
\begin{equation}
P(x) = \frac{2}{2 \zeta^3-4 \zeta^2+1} \left[ (1-2 \zeta^2)-(1-2 \zeta) x
\right]
\end{equation}
for $\rho$ and $a_1$ mesons, where $\zeta = m_{\rho,a_1}^2/m_\tau^2$. We assume
a $\tau$ jet tagging efficiency of 50\%. In certain measurements the use of $c$
tagging is necessary as well, and we assume a 50\% efficiency, the same one that
it is expected at ILC \cite{ctag}.

We simulate the calorimeter and tracking resolution of the detector by
performing a Gaussian smearing of the energies of electrons ($e$), muons ($\mu$)
and jets ($j$), using the expected resolutions \cite{clic04},
\begin{equation}
\frac{\Delta E^e}{E^e} = \frac{10\%}{\sqrt{E^e}} \oplus 1 \% \;, \quad
\frac{\Delta E^\mu}{E^\mu} = 0.005 \% \, E^\mu \;, \quad
\frac{\Delta E^j}{E^j} = \frac{50\%}{\sqrt{E^j}} \oplus 4 \% \;,
\end{equation}
where the two terms are added in quadrature and the energies are in GeV.
We apply kinematical cuts on transverse momenta, $p_T \geq 10$ GeV, and
pseudorapidities $|\eta| \leq 2.5$, the latter corresponding to polar angles
$10^\circ \leq \theta \leq 170^\circ$. We reject events in which the
leptons or jets are not isolated, requiring a ``lego-plot'' separation
$\Delta R = \sqrt{\Delta \eta^2+\Delta \phi^2} \geq 0.4$.
For the Monte Carlo integration in 6-body phase space we use
{\tt RAMBO} \cite{rambo}.

In electron and muon final states the light neutrino momentum $p_\nu$ is
determined from the missing transverse and longitudinal momentum of the event
and the requirement that $p_\nu^2 = 0$ 
(despite ISR and beamstrahlung, the missing longitudinal momentum approximates
with a reasonable accuracy the original neutrino momentum).
In final states with $\tau$
leptons, the reconstruction is more involved, due to the secondary neutrino
from the $\tau$ decay. We determine the ``primary'' neutrino momentum and the
fraction $x$ of the $\tau$ momentum retained by the $\tau$ jet using the
kinematical constraints
\begin{align}
E_W + E_\nu + \frac{1}{x} E_j & = \sqrt s \,, \nonumber \\
\vec p_W + \vec p_\nu + \frac{1}{x}  \vec p_j & = 0 \,, \nonumber \\
p_\nu^2 & = 0 \,,
\end{align}
in obvious notation.
These constraints only hold if ISR and beamstrahlung are ignored, and in the
limit of perfect detector resolution. When solving them for the generated Monte
Carlo events we sometimes obtain $x > 1$ or $x < 0$. In the first case we
arbitrarily set $x=1$, and in the second case we set $x = 0.55$, which is the
average momentum fraction of the $\tau$ jets. With the procedure outlined here,
the reconstructed $\tau$ momentum reproduces with a fair accuracy the original
one, while the $p_\nu$ obtained is often completely different from its actual
value.

\section{Heavy neutrino discovery at CLIC and determination of their properties}
\label{sec:5}

We address in turn the discovery of a new heavy neutrino (sections \ref{sec:5.1}
and \ref{sec:5.2}), the determination of its Dirac or Majorana character
(section \ref{sec:5.3}) and the
measurement of its couplings to $e$, $\mu$, $\tau$ (section \ref{sec:5.4}). We
try to be as concise as possible without losing generality or omitting the main
points.
Following the discussion in section \ref{sec:3} we can distinguish two
interesting scenarios for our analysis:
({\em i\/}) the heavy neutrino only mixes with the electron; ({\em ii\/}) it
mixes with $e$ and either $\mu$, $\tau$, or both. For the study of the $m_N$
dependence and the determination of the neutrino nature
we assume for simplicity that $N$  only mixes with the electron. Additionally,
we assume that the neutrino is a
Majorana fermion in sections \ref{sec:5.1}, \ref{sec:5.2} and \ref{sec:5.4},
where the results obtained are almost independent of its character.

\subsection{Discovery of a heavy neutrino}
\label{sec:5.1}

\begin{figure}[htb]
\begin{center}
\epsfig{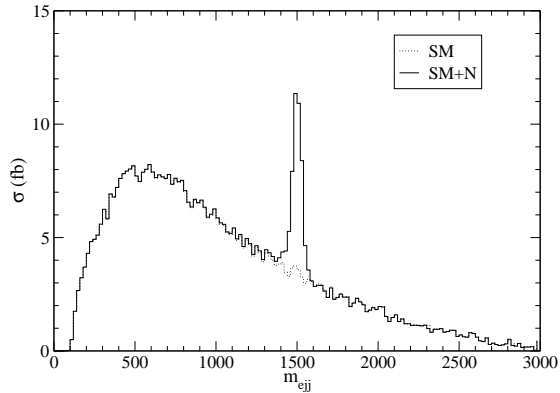}
\caption{Kinematical distribution of the $ejj$ invariant mass, for $m_N = 1500$
GeV.}
\label{fig:mn}
\end{center}
\end{figure}

The existence of a heavy neutrino which couples to the electron can be detected
as a
sharp peak in the distribution of the $ejj$ invariant mass $m_{ejj}$, plotted in
Fig.~\ref{fig:mn} for $V_{eN} = 0.05$. The dotted and solid lines correspond to
the SM and SM plus a 1500 GeV Majorana neutrino, respectively. For a
Dirac neutrino the results do not change. The width of the peak is mainly due
to energy smearing included in our Monte Carlo, and the intrinsic $N$
width, $\Gamma_N = 8.2$ GeV and $\Gamma_N = 4.1$ GeV for a Majorana and Dirac
neutrino, respectively, has a smaller influence in this case.

The SM and SM plus heavy neutrino cross sections are
collected in Table~\ref{tab:cs}, before and after the kinematical cut
\begin{equation}
1460 ~\mathrm{GeV} \leq m_{ejj} \leq 1540 ~\mathrm{GeV} \,.
\label{ec:cut}
\end{equation}
The criterion used here for the discovery of the new neutrino is that the excess
of events\footnote{It must be stressed that the SM cross section at the peak can
be calculated and normalised using the measurements far from this region.} (the
signal $S$) in the peak region defined by Eq.~(\ref{ec:cut}) amounts to more
than 5 standard deviations of the number of expected events (the background
$B$), that is, $S/\sqrt B \geq 5$.
This ratio is larger than $5$ for $V_{eN} \geq 7.8 \times 10^{-3}$,
which is the minimum mixing angle for which a 1500 GeV neutrino can be
discovered.
Conversely, if no signal is found, the limit $V_{eN} \leq 4.5 \times
10^{-3}$ can be set at 90\% confidence level (CL), improving the present limit
$V_{eN} \leq 0.073$ by a factor of 16.

\begin{table}[htb]
\begin{center}
\begin{tabular}{lcc}
& No cut & With cut \\
\hline
SM       & 516 & 14.6 \\
SM + $N$ & 548 & 39.4 
\end{tabular}
\caption{Cross sections (in fb) for $e^+ e^- \to e^\mp W^\pm \nu$ before and
after the kinematical cut in Eq.~(\ref{ec:cut}).}
\label{tab:cs}
\end{center}
\end{table}

In the most general case that $N$ simultaneously mixes with the
three charged leptons, there may be in principle signals in the $e$, $\mu$ and
$\tau$ channels, and the three of them must be experimentally analysed.
We choose equal values $V_{eN} = V_{\mu N} = V_{\tau N} = 0.04$ to illustrate
the relative sensitivities of the three channels. For electron and muon final
states we apply the kinematical cut in Eq.~(\ref{ec:cut}), while for
taus the distribution is broader and we use
\begin{equation}
1420 ~\mathrm{GeV} \leq m_{\tau jj} \leq 1580 ~\mathrm{GeV} \,.
\label{ec:cut2}
\end{equation}
The SM and SM plus heavy neutrino cross sections after these kinematical
cuts\footnote{For
the muon and tau channels the background is dominated by resonant $W^+ W^-$
production, thus a cut on the $m_{\ell \nu}$ invariant mass could
reduce it significantly. However, in practice it may be very difficult to
reconstruct the $W$ mass at CLIC energy, and to be conservative we do not apply
any cut on $m_{\ell \nu}$.}
can be found in Table~\ref{tab:cs2}.
For these values of the couplings, the heavy neutrino signal could be seen with
a statistical significance of
$\sim 40 \sigma$, $\sim 250 \sigma$ and $\sim 70 \sigma$ 
in the $e$, $\mu$ and $\tau$ channels, respectively, after one
year of running. 

\begin{table}[htb]
\begin{center}
\begin{tabular}{lccc}
& $e$  & $\mu$ & $\tau$ \\
\hline
SM        & 14.6 & 0.36  & 0.096 \\
SM + $N$  & 19.5 & 5.24  & 1.19 
\end{tabular}
\caption{Cross sections (in fb) for $e^+ e^- \to \ell^\mp W^\pm \nu$, for
$\ell=e,\mu,\tau$, including the kinematical cuts on $m_{\ell jj}$.}
\label{tab:cs2}
\end{center}
\end{table}

\begin{figure}[htb]
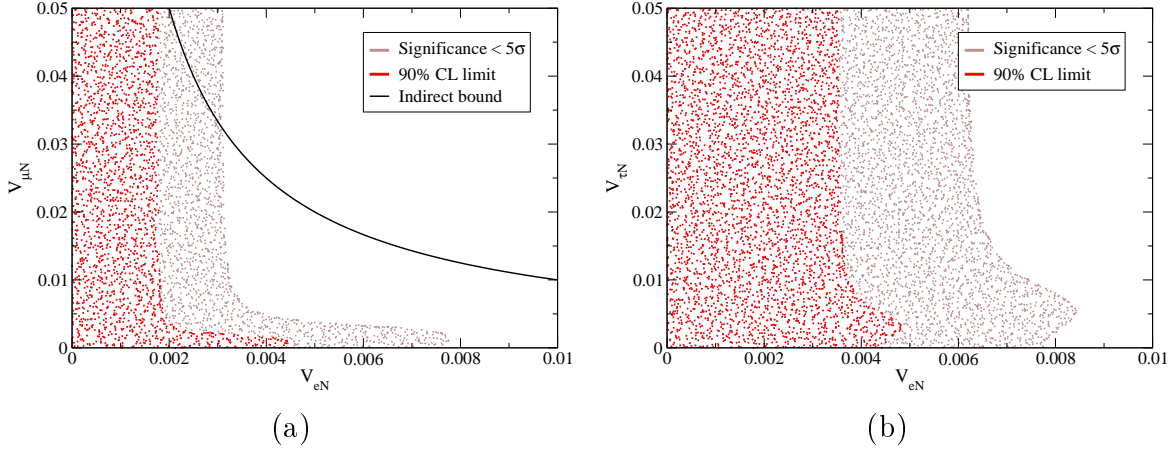

\begin{center}
\begin{tabular}{cc}
\epsfig{file=Figs/bound-em.eps,height=5.2cm,clip=} &
\epsfig{file=Figs/bound-et.eps,height=5.2cm,clip=} \\
(a) & (b)
\end{tabular}
\caption{Combined limits on: $V_{eN}$ and $V_{\mu N}$, for $V_{\tau N} = 0$ (a);
$V_{eN}$ and $V_{\tau N}$, for $V_{\mu N} = 0$ (b), for $m_N = 1500$ GeV. The
red areas represent the 90\% CL limits if no signal is observed. The white areas
extend up to present bounds $V_{eN} \leq 0.073$, $V_{\mu N} \leq 0.098$,
$V_{\tau N} \leq 0.13$, and correspond to the region where a combined
statistical significance of $5\sigma$ or larger is achieved. The indirect limit
from $\mu - e$ LFV processes is also shown.}
\label{fig:limits}
\end{center}
\end{figure}

We clearly see that the muon channel is much more sensitive for equal values of
the couplings. An $eNW$ interaction is absolutely necessary to produce the heavy
neutrino at observable rates but, once it has been produced, its decays in the
muon and tau channels are easier to spot over the background. Since the
production mechanism is
strongly dominated by $t$ and $u$-channel $W$ exchange diagrams (see section
\ref{sec:3}), the observed signals $S_e$, $S_\mu$, $S_\tau$ can be written as
\begin{equation}
S_\ell = A_\ell \, V_{eN}^2 \;
\frac{V_{\ell N}^2}{V_{eN}^2 + V_{\mu N}^2 + V_{\tau N}^2}
\label{ec:13}
\end{equation}
to an excellent approximation. The common factor $V_{eN}^2$ comes from the
production, the ratio of couplings corresponds to the decay and $A_\ell$ are
constants. Using
the data in Table \ref{tab:cs2} and Eqs. (\ref{ec:13}) we can obtain the
combined limits on $V_{eN}$ and $V_{\mu N}$ or $V_{\tau N}$ plotted in
Fig.~\ref{fig:limits}. It is very interesting to observe that a small coupling
to the muon $V_{\mu N} \gtrsim 0.005$ greatly increases the sensitivity to
$V_{eN}$, from $\sim 0.008$ to $\sim 0.0035$, due to the better observation of
the heavy neutrino in the muon channel. For a fixed $V_{eN}$ and increasing
$V_{\mu N}$, the $eW\nu$ channel becomes less significant because of the smaller
branching ratio, and the observation is better in the $\mu W \nu$ channel.
(For ILC the behaviour is rather different, see next section.)
In the case of the tau the effect is similar but less pronounced. 

For comparison, we include in Fig.~\ref{fig:limits} (a) the indirect limit
on $V_{eN}$, $V_{\mu N}$ derived from low energy LFV
processes. For $V_{\mu N}$ smaller than 0.05, the direct limit obtained by the
absence of heavy neutrino production is much better than the indirect one.
Moreover, the latter can be evaded
if we allow for cancellations between heavy neutrino contributions, as discussed
in section~\ref{sec:2} (see also Ref.~\cite{paper1}).
In the case of the tau lepton, the indirect limit from
LFV processes is less stringent than the direct limits $V_{eN} \leq 0.073$,
$V_{\tau N} \leq 0.13$ and is not shown.

When the heavy neutrino does not couple to the electron but only to the muon
and/or tau, the only Feynman diagram contributing to the signal is
\ref{fig:diagN2}c. For CLIC the situation is much worse than for ILC
\cite{paper1} because at a higher CM energy this diagram is more suppressed.
Heavy neutrino production rates are thus negligibly small, giving an excess of a
handful of events (for an integrated luminosity of 1000 fb$^{-1}$) over the SM
background even for $V_{\mu N}$, $V_{\tau N}$ in their upper experimental
bounds.

\subsection{Dependence on the heavy neutrino mass}
\label{sec:5.2}

\begin{figure}[htb]
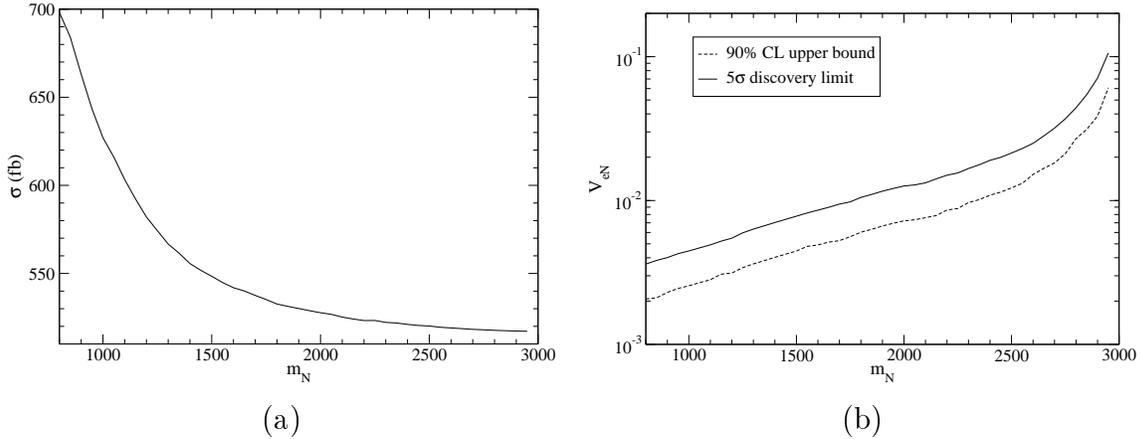

\begin{center}
\begin{tabular}{cc}
\epsfig{file=Figs/mass-cross.eps,width=7.3cm,clip=} & 
\epsfig{file=Figs/mass-coup.eps,width=7.3cm,clip=} \\
(a) & (b)
\end{tabular}
\caption{(a) Cross section for $e^+ e^- \to e^\mp W^\pm \nu$ for $V_{eN} =
0.05$ and different values of $m_N$.
(b) Dependence of the discovery and upper limits on $V_{eN}$ on the
heavy neutrino mass. Both plots assume mixing only with the electron.}
\label{fig:massdep}
\end{center}
\end{figure}

The cross section for $e^+ e^- \to e^\mp W^\pm \nu$ including the heavy neutrino
contribution
exhibits a moderate dependence on $m_N$, which is mainly due to phase space.
The variation of the total $e^\mp W^\pm \nu$ cross section (including ISR,
beamstrahlung
and ``detector cuts'' as explained in section~\ref{sec:4}) with $m_N$
can be seen in Fig.~\ref{fig:massdep} (a). For larger masses the
cross sections are smaller an thus the limits on $V_{eN}$ are worse. This
behaviour is attenuated by the fact that the SM background also decreases
for larger $m_{ejj}$, as can be clearly observed in
Fig.~\ref{fig:mn}. The resulting limits on the heavy neutrino coupling
are shown in Fig.~\ref{fig:massdep} (b) as a function of $m_N$. These limits
assume implicitly that the heavy neutrino only mixes with the electron (for
mixing also with the muon they improve, as seen in the previous subsection). The
kinematical cuts are not optimised for each value of $m_N$.

\subsection{Determination of the Majorana or Dirac nature}
\label{sec:5.3}

The cross section for $e^- W^+ \nu$ production mediated by a heavy neutrino
is fairly insensitive to its Dirac or Majorana character. Still, the different
production mechanisms show up in the angular distribution of $N$ with respect to
the incoming electron.  
In the case of a heavy Majorana neutrino the new contribution is dominated by
diagrams \ref{fig:diagN}a, \ref{fig:diagN}b, leading to a forward-backward
symmetric distribution for
the production angle $\varphi_{Ne^-}$ between $N$ and the incoming electron.
The
distribution peaks at $\cos \varphi_{Ne^-} = 1$ when
$t \equiv (p_N - p_{e^-})^2 = 0$
and the first diagram is enhanced, and at $\cos \varphi_{Ne^-} = -1$ when
$u \equiv (p_N - p_{e^+})^2 = 0$ (and the second one is enhanced).

\begin{figure}[htb]
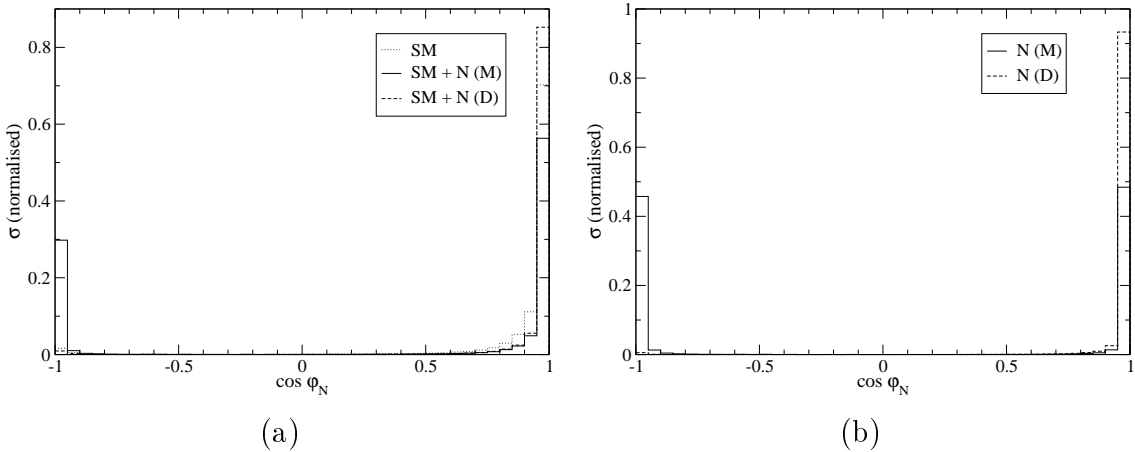

\begin{center}
\begin{tabular}{cc}
\epsfig{file=Figs/cosph-N.eps,width=7.3cm,clip=} &
\epsfig{file=Figs/cosph-Nr.eps,width=7.3cm,clip=} \\
(a) & (b)
\end{tabular}
\caption{(a) Dependence of the cross section on the angle $\varphi_N$, for the
SM and the SM plus a 1500 GeV Majorana (M) or Dirac (D) neutrino. (b) The same,
but subtracting the SM contribution.}
\label{fig:cosph-N}
\end{center}
\end{figure}

In the case of Dirac neutrinos the $u$-channel diagram is absent, and the
distribution only peaks at $\cos \varphi_{Ne^-} = 1$ for final states with
$e^-,\mu^-,\tau^-$ and at $\cos \varphi_{Ne^-} = -1$ for the charge conjugate
processes with $e^+,\mu^+,\tau^+$. It is then convenient to define
$\varphi_N \equiv \varphi_{Ne^-}, \varphi_{Ne^+}$ for $e^- W^+ \nu$,
$e^+ W^- \nu$ events, respectively. Its normalised distribution is shown
in Fig.~\ref{fig:cosph-N} (a) for events surviving the kinematical cut in
Eq.~(\ref{ec:cut}). We consider the SM, and SM plus a
Majorana or Dirac neutrino. The most conspicuous results are obtained
subtracting the SM contribution, which can be calculated and calibrated using
the measurements outside the peak. In this case, Fig.~\ref{fig:cosph-N} (b), 
the difference between the Dirac and Majorana cases is apparent.
We remark that the signal cross
section at the peak is of 24.8 fb for
$V_{eN} = 0.05$, what gives a sufficiently large event sample to
distinguish both cases even for smaller mixing angles (see also
Fig.~\ref{fig:massdep}).

\begin{figure}[htb]
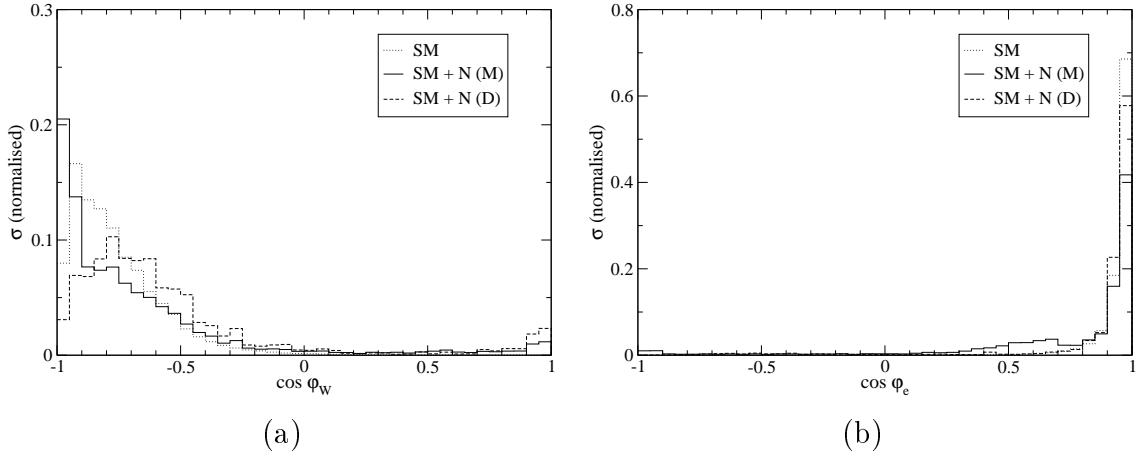

\begin{center}
\begin{tabular}{cc}
\epsfig{file=Figs/cosph-W.eps,width=7.3cm,clip=} &
\epsfig{file=Figs/cosph-e.eps,width=7.3cm,clip=} \\
(a) & (b)
\end{tabular}
\caption{Dependence of the cross section on the angles $\varphi_W$ (a) and
$\varphi_e$ (b), for the SM and the SM plus a 1500 GeV Majorana or Dirac
neutrino.}
\label{fig:cosph-We}
\end{center}
\end{figure}

For completeness, we also show in Fig.~\ref{fig:cosph-We} the dependence of the
cross section on the angles $\varphi_W$, $\varphi_e$ between 
the produced $W^\pm$, $e^\mp$ and the
incoming electron (for $e^-$ final states) or positron (for final $e^+$). We
 restrict
ourselves to events surviving the kinematical cut in Eq.~(\ref{ec:cut}), as
in the previous case. Although these two distributions also show some
sensitivity to the Dirac or Majorana character of the neutrino, it is obvious
when compared to Fig.~\ref{fig:cosph-N} (a)
that the best results are obtained from the analysis of the polar angle of the
produced neutrino $\varphi_N$.

\subsection{Measurement of heavy neutrino couplings}
\label{sec:5.4}

In order to measure the moduli of the heavy neutrino couplings to charged
leptons, the constants $A_\ell$ in Eq.~(\ref{ec:13}) 
must be theoretically calculated. This can be
done in principle
with full Monte Carlo simulations taking into account all radiation and
hadronisation effects, as well as the real detector behaviour. The couplings of
the heavy neutrino are then
\begin{eqnarray}
V_{eN}^2 & = & \frac{S_e}{A_e} + \frac{S_\mu}{A_\mu} + 
\frac{S_\tau}{A_\tau} \,, \nonumber \\
\frac{V_{\ell N}^2}{V_{eN}^2} & = & \frac{S_\ell}{A_\ell}
\left( \frac{S_e}{A_e} \right)^{-1} \,,\quad \quad \ell=\mu,\tau \,.
\end{eqnarray}
The uncertainty in their measurement comes from the statistical fluctuations of
the signal and background, as well as from the theoretical calculation of the
constants $A_\ell$. Among other factors, their calculation
is affected by the inherent uncertainties in the signal normalisation. We expect
that $A_\ell$ can be obtained with a precision
of 10\%, which in principle does not affect the determination of coupling
ratios.

We estimate the accuracy with which heavy neutrino couplings could be extracted,
calculating the $A_\ell$ constants from the peak cross sections given by our
Monte Carlo for a ``reference'' set of couplings, and assuming a common 10\%
error in their determination. Then, using
as input the cross sections for  $V_{eN} = V_{\mu N} = V_{\tau N} = 0.04$, the
couplings obtained are
\begin{eqnarray}
V_{eN} & = & 0.0388 \pm 0.00034 ~\mathrm{(stat)} \pm 0.0019 ~\mathrm{(sys)} \,,
 \nonumber \\ 
V_{\mu N}/V_{eN} & = & 1.007 \pm 0.016 ~\mathrm{(stat)} \,, \nonumber \\
V_{\tau N}/V_{eN} & = & 1.030 \pm 0.028 ~\mathrm{(stat)} \,.
\end{eqnarray}
In this case the statistical precision of the ratios is very good, of
a $2-3$\%. The uncertainty in $V_{eN}$ is of a 5\%, dominated by systematics.

A second issue is the determination of the chirality of $\ell N W$ couplings.
This can be done with the measurement of the $\ell s$ forward-backward (FB)
asymmetry in the $W$ rest frame \cite{lampe,PRD}, which is sensitive to the
chiral structure of the $\ell N W$ coupling involved in the decay
$N \to \ell^- W^+ \to \ell^- c \bar s$ (and in its charge conjugate). The
measurement of this asymmetry requires to distinguish between the two quark jets
resulting from $W^+$ decay, what can be done restricting ourselves to $W^+ \to
c \bar s$ and taking advantage of $c$ tagging to require a tagged $c$ jet in the
final state. This reduces the cross sections by a factor of four.

We define $\theta_{\ell s}$ as the angle between the momenta of the
charged lepton $\ell$ and the $\bar s$ jet, in the $W$ rest frame (the
definition is the same for $N$ decays into $\ell^- W^+ \to \ell^- c \bar s$ and
$\ell^+ W^- \to \ell^+ \bar c s$). The FB asymmetry is defined as
\begin{equation}
A_\mathrm{FB} = \frac{N(\cos \theta_{\ell s} > 0) - N(\cos \theta_{\ell s}
< 0)}{N(\cos \theta_{\ell s} > 0) + N(\cos \theta_{\ell s} < 0)} \,,
\end{equation}
with $N$ standing for the number of events. If we parameterise a general
$\ell N W$ vertex (ignoring effective $\sigma^{\mu \nu}$ terms) as
\begin{equation}
\mathcal{L}_{\ell WN} = - \frac{g}{\sqrt 2} \, \bar \ell \gamma^\mu \left(
 g_L P_L + g_R P_R \right) N \; W_\mu + \mathrm{H.c.} \,
\end{equation}
and assume that the $W$ coupling to quarks is purely left-handed, the FB
asymmetry is
\begin{equation}
A_\mathrm{FB} = \frac{3 M_W^2}{4 M_W^2+2 m_N^2} \,
\frac{|g_L|^2-|g_R|^2}{|g_L|^2+|g_R|^2} \,.
\label{ec:AFB}
\end{equation}
In the case of heavy neutrino singlets $g_L = V_{\ell N}$, $g_R = 0$ as seen in
Eq.~(\ref{ec:lNW}), and the second
factor in Eq.~(\ref{ec:AFB}) equals unity. Still, the first factor is very small
for $m_N \gg M_W$, and for $m_N = 1500$ GeV we have
$A_\mathrm{FB} = 4.3 \times 10^{-3}$ for the three lepton flavours. Such
asymmetries are unobservable, as long as the experimental statistical errors
expected in the experiment
are typically $\Delta A_\mathrm{FB} \gtrsim 0.012$.\footnote{In the absence of
background, the
statistical error of an asymmetry $A$ is given by the simple expression
$\Delta A = \sqrt{(1-A^2)/(S \, L)}$, with $S$ the signal cross section and $L$
the integrated luminosity. With an asymmetric background the expression is more
involved but it can be seen that $\Delta A$ is larger than without background.}
Still, for $m_N$ of several hundreds of GeV the asymmetries are measurable.
We analyse such possibility in next section.

Further observables are sensitive to the structure of the $\ell NW$ vertices,
namely spin asymmetries. A proper analysis requires the search for a direction
along which the $N$ polarisation is maximal, what is beyond the scope of this
work. Besides, the information on couplings extracted from such observables is
expected to be less clean, since they involve additional variables apart from
masses and couplings.

\section{Heavy non-decoupled neutrinos at ILC}
\label{sec:6}

Heavy neutrinos with masses up to a few hundreds of GeV can also be produced at
ILC
with a CM energy of 500 GeV. Many features of the production process are common
to both mass and CM energy scales, but in some other respects ILC and CLIC are
rather
different. In order to have a better comparison between both cases we summarise
here several results for ILC extending the work in Ref.~\cite{paper1}.
We take a mass $m_N = 300$ GeV for most of our computations,
which follow closely what is done for CLIC (see also Ref.~\cite{paper1} for
details). The integrated luminosity is assumed to be 345 fb$^{-1}$,
corresponding to one year of running.

\begin{figure}[htb]
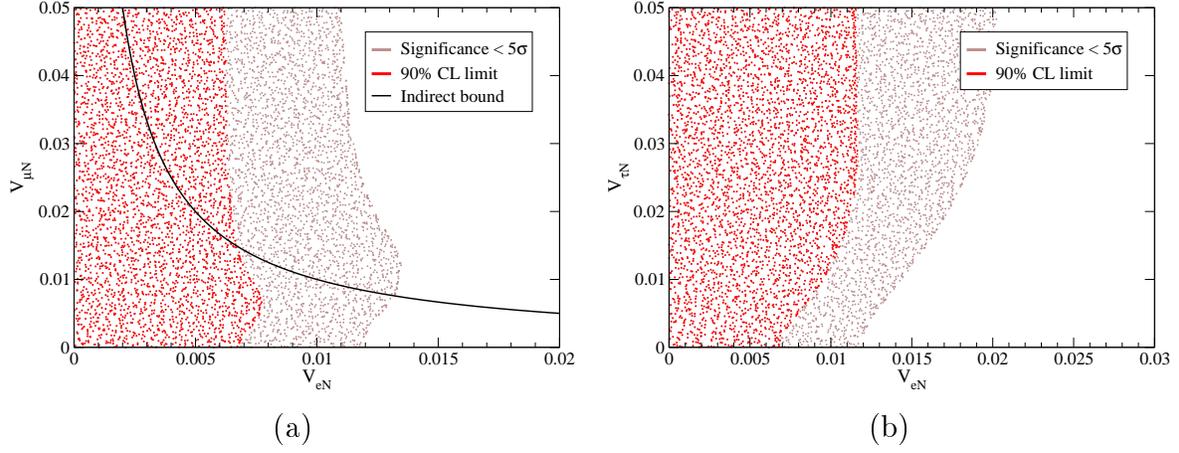

\begin{center}
\begin{tabular}{cc}
\epsfig{file=Figs-ILC/bound-em.eps,height=5.2cm,clip=} &
\epsfig{file=Figs-ILC/bound-et.eps,height=5.2cm,clip=} \\
(a) & (b)
\end{tabular}
\caption{Combined limits obtained at ILC on: $V_{eN}$ and $V_{\mu N}$, for
$V_{\tau N} = 0$ (a); $V_{eN}$ and $V_{\tau N}$, for $V_{\mu N} = 0$ (b). The
red areas represent the 90\% CL limits if no signal is observed. The white areas
extend up to present bounds $V_{eN} \leq 0.073$, $V_{\mu N} \leq 0.098$,
$V_{\tau N} \leq 0.13$, and correspond to the region where a combined
statistical significance of $5\sigma$ or larger is achieved. The indirect limit
from $\mu - e$ LFV processes is also shown. We take $m_N = 300$ GeV.}
\label{fig:limits-ILC}
\end{center}
\end{figure}

With the same signal reconstruction method as for CLIC we obtain for $m_N = 300$
GeV the combined limits on $V_{eN}$ and $V_{\mu N}$ or $V_{\tau N}$ plotted in
Fig.~\ref{fig:limits-ILC}.
In contrast with the behaviour obtained at CLIC, at ILC the sensitivities in the
muon and electron channels are similar, and both are better than in
$\tau W \nu$
production. This can be clearly observed in both plots: a $\mu N W$ coupling has
little effect on the limits on $V_{eN}$, but a coupling with the tau decreases
the sensitivity, because the decays in the tau channel are harder to observe.
The direct limit on $V_{eN}$, $V_{\mu N}$ obtained here improves the indirect
one only for $V_{\mu N} \lesssim 0.01$. However, as it has already been
remarked, the latter is not general and can be evaded with cancellations among
heavy neutrino contributions \cite{paper1}.

\begin{figure}[t]
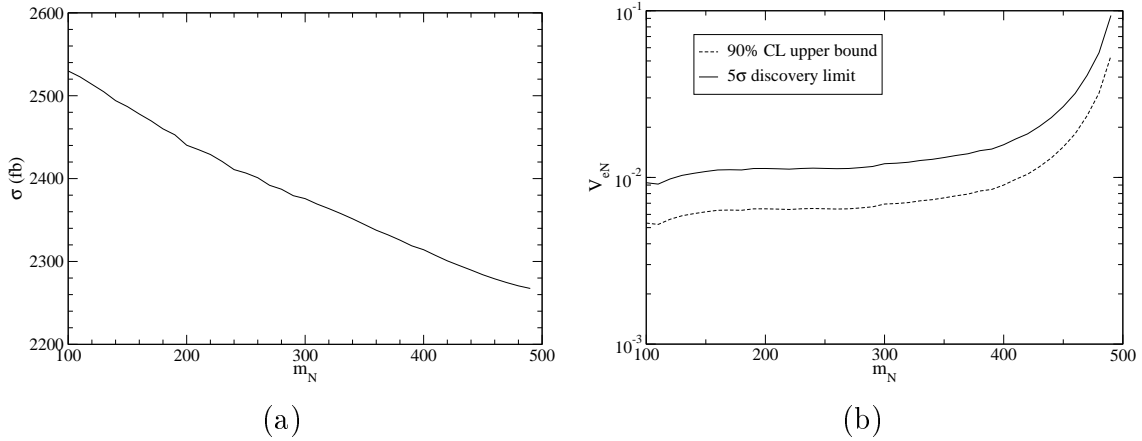

\begin{center}
\begin{tabular}{cc}
\epsfig{file=Figs-ILC/mass-cross.eps,width=7.3cm,clip=} & 
\epsfig{file=Figs-ILC/mass-coup.eps,width=7.3cm,clip=} \\
(a) & (b)
\end{tabular}
\caption{(a) Cross section for $e^+ e^- \to e^\mp W^\pm \nu$ at ILC for
$V_{eN} = 0.073$ and different values of $m_N$.
(b) Dependence of the discovery and upper limits on $V_{eN}$ on the
heavy neutrino mass. Both plots assume mixing only with the electron.}
\label{fig:massdep-ILC}
\end{center}
\end{figure}

The dependence of the total $e^\mp W^\pm \nu$ cross section on $m_N$
can be seen in Fig.~\ref{fig:massdep-ILC} (a), for $V_{eN} = 0.073$, $V_{\mu N}
= V_{\tau N} = 0$. For a heavier $N$
the cross sections are smaller and thus the limits on $V_{eN}$ are worse.
However, up to $m_N = 400$ GeV this is compensated by the fact that the SM
background also decreases for larger $m_{ejj}$. The limits on $V_{eN}$ are
shown in Fig.~\ref{fig:massdep-ILC} (b) as a function of $m_N$, assuming
that the heavy neutrino only mixes with the electron.

\begin{figure}[htb]
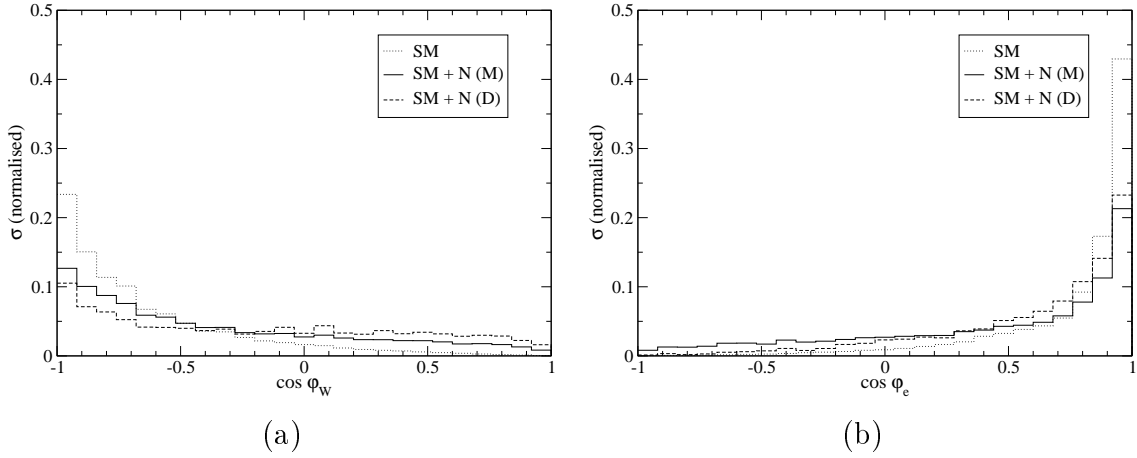

\begin{center}
\begin{tabular}{cc}
\epsfig{file=Figs-ILC/cosph-W.eps,width=7.3cm,clip=} &
\epsfig{file=Figs-ILC/cosph-e.eps,width=7.3cm,clip=} \\
(a) & (b)
\end{tabular}
\caption{Dependence of the cross section at ILC on the angles $\varphi_W$ (a)
and $\varphi_e$ (b), for the SM and the SM plus a 300 GeV Majorana or Dirac
neutrino.}
\label{fig:cosph-We-ILC}
\end{center}
\end{figure}

If a heavy neutrino is discovered at ILC, its nature can be uniquely determined
with the study of the $\varphi_N$ dependence of the cross section, as it has
been discussed
in section \ref{sec:5.3}. The dependence of the peak cross section on
$\varphi_N$ is shown in Fig.~\ref{fig:cosph-N-ILC} (a). The results after
subtracting the SM contribution can be seen in Fig.~\ref{fig:cosph-N-ILC} (b). 
The distribution is less concentrated at $\cos \varphi_N = \pm 1$ than for CLIC
energy but allows to determine unambiguously the neutrino character even
if a relatively small number of signal events is collected. The dependence of
the peak cross section on $\varphi_W$ and $\varphi_e$ is presented in
Fig.~\ref{fig:cosph-We-ILC}.\footnote{It should be noted that the $\varphi_W$
and $\varphi_e$ distributions presented in Ref.~\cite{paper1} correspond to the
whole range of $m_{ejj}$ and not only to the peak.}

\begin{figure}[htb]
\begin{center}
\begin{tabular}{cc}
\epsfig{file=Figs-ILC/cosph-N.eps,width=7.3cm,clip=} &
\epsfig{file=Figs-ILC/cosph-Nr.eps,width=7.3cm,clip=} \\
(a) & (b)
\end{tabular}
\caption{(a) Dependence of the cross section at ILC on the angle $\varphi_N$,
for the
SM and the SM plus a 300 GeV Majorana (M) or Dirac (D) neutrino. (b) The same,
but subtracting the SM contribution.}
\label{fig:cosph-N-ILC}
\end{center}
\end{figure}

\begin{figure}[htb]
\begin{center}
\begin{tabular}{cc}
\epsfig{file=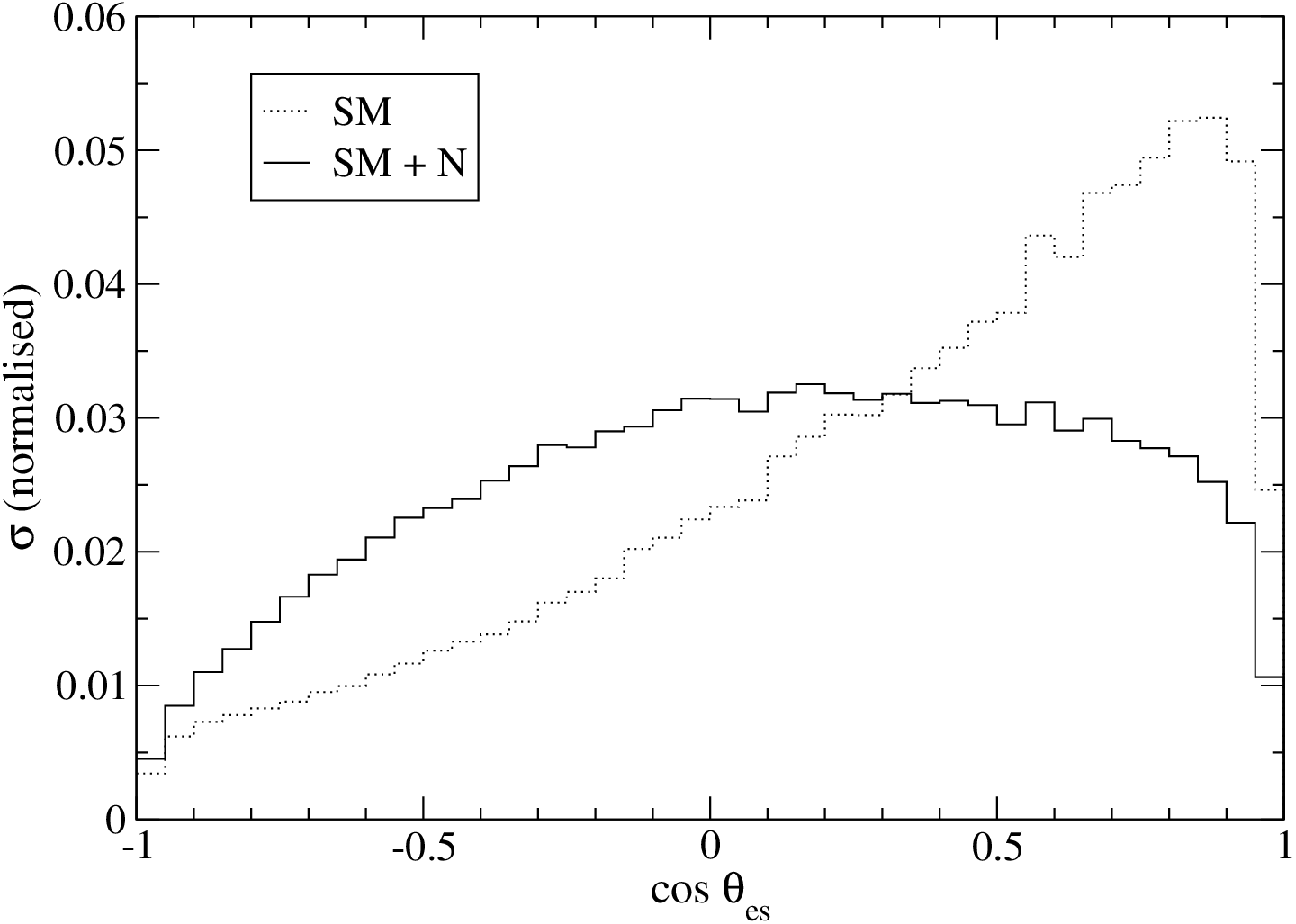,width=7.3cm,clip=} &
\epsfig{file=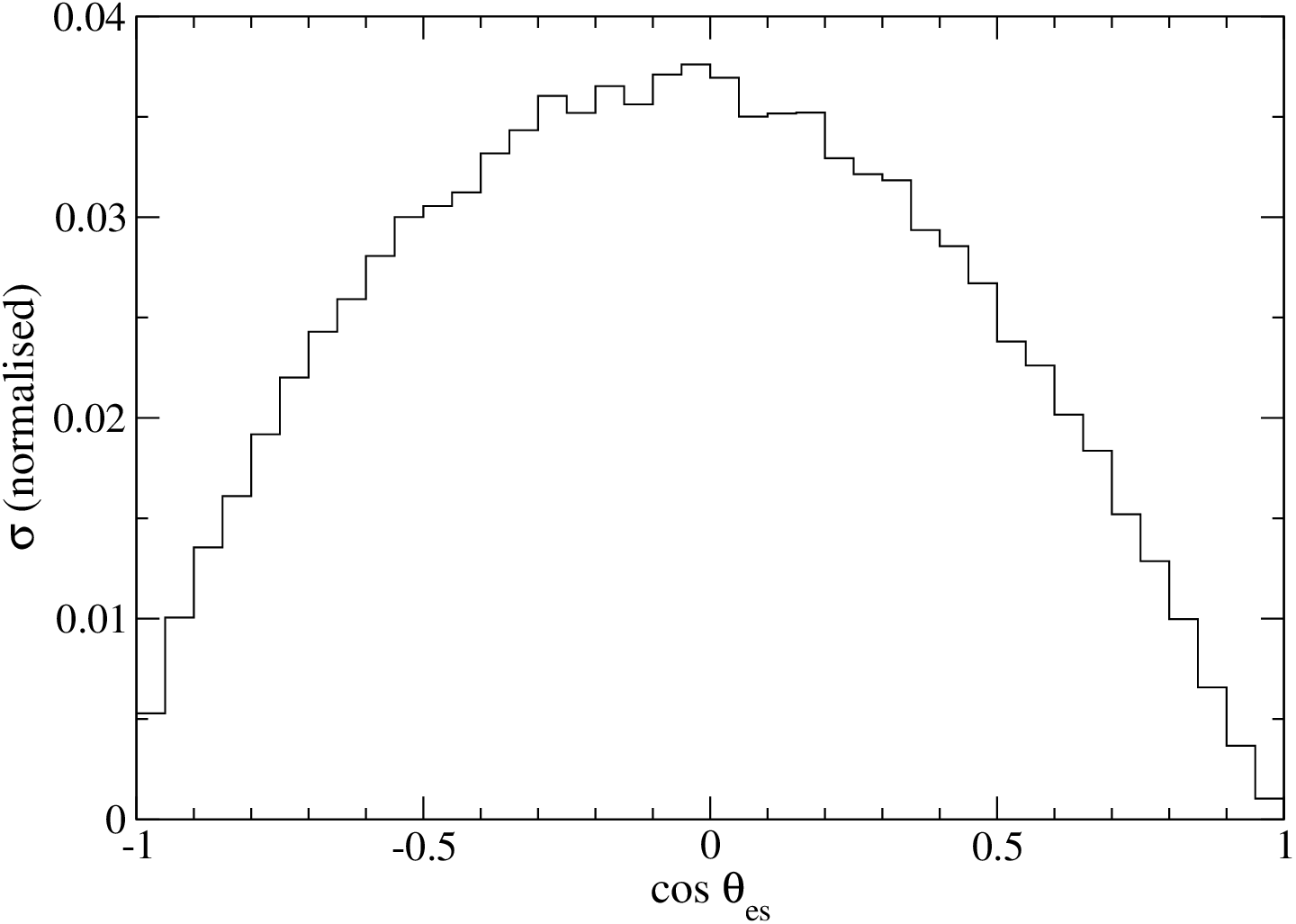,width=7.3cm,clip=} \\
(a) & (b)
\end{tabular}
\caption{(a) Dependence of the cross section at ILC on the angle $\theta_{e
s}$ for the SM and the SM plus a 300 GeV Majorana neutrino. (b) The same,
but subtracting the SM contribution.}
\label{fig:costh-FB-ILC}
\end{center}
\end{figure}

The measurement at ILC of the $V_{eN}$, $V_{\mu N}$, $V_{\tau N}$ moduli
can be done in the
same way described for CLIC, without any substantial differences, thus we do not
repeat that discussion here. Still, an important difference appears in the
determination of their chiral structure. In contrast with TeV scale neutrinos,
for $m_N = 300$ GeV the FB asymmetry is observable. We show in
Fig.~\ref{fig:costh-FB-ILC} (a) the dependence of the SM
and SM plus heavy neutrino peak cross sections with respect to the angle
$\theta_{es}$. The measurement of the FB asymmetry requires the subtraction of
the SM prediction, giving the distributions shown in Fig.~\ref{fig:costh-FB-ILC}
(b). For a Majorana neutrino with $V_{eN} = 0.073$,
$V_{\mu N} = V_{\tau N} = 0$, the resulting asymmetry
is $A_\mathrm{FB} = 0.083 \pm 0.016$, where the error quoted is statistical.
The SM and SM$+N$ cross sections after $c$ tagging are 13.4 and 32.7 fb,
including only $W$ hadronic decays with a $c$ quark.
For a Dirac neutrino the results found are equivalent within Monte Carlo
uncertainty. The FB asymmetry obtained from the simulation is in good agreement
with the
theoretical value $A_\mathrm{FB} = 0.094$ expected for a purely left-handed
coupling.

If a heavy neutrino signal is observed in this mass range, we expect that
the FB asymmetry will have a sufficient statistical
significance so as to indicate that the $eNW$ coupling is left-handed, at least
after several years of running. The precision of the $A_\mathrm{FB}$ measurement
strongly depends on the size of the $N$ cross section, therefore the
possibility of extracting the left- and right-handed parts of the coupling
from the total cross section and $A_\mathrm{FB}$ measurements is difficult to
assess in general.

\section{Conclusions}
\label{sec:7}

Neutrinos with masses of the order of 1 TeV are predicted by many models
attempting to make the new physics scale observable at future colliders.
Their presence leads to a greater complexity of the models (in order to
reproduce the light neutrino masses),  especially when they are not decoupled
from the light fermions. Nevertheless, if they exist, large $e^+e^-$ colliders
can discover them or provide the best limits on their masses and mixings. 
Such discovery would give a new insight into the mechanism for
neutrino mass generation.

In this paper we have concentrated on the production of heavy neutrino singlets
in association with a light neutrino. If they are the only addition to the SM,
the production cross section for $NN$ pairs is
suppressed by extra mixing angle factors $V_{\ell N}^2$. This is also reduced by
the smaller
phase space and by additional decay branching ratios, what makes this process
much less sensitive to the presence of these heavy fermions.\footnote{For
$e^+ e^- \to NN$ the $t$
and $u$-channel exchange diagrams are quadratic in $V_{eN}$, and the $s$-channel
$Z$ diagram has the mixing factor $X_{NN}$ in Eq.~(\ref{ec:lz}), which is equal
to $|V_{eN}|^2+|V_{\mu N}|^2+|V_{\tau N}|^2$. Therefore, the cross section
is proportional to light-heavy mixing angles to the fourth power. For $m_N = 1$
TeV and a coupling $V_{eN} = 0.004$ in the discovery limit, the extra mixing
angle suppression 
of this cross section is of order $V_{eN}^2 = 1.6 \times 10^{-5}$, already
giving unobservable rates for heavy neutrino pair production.}
In left-right models the new gauge interactions allow to
produce heavy neutrinos in pairs, or in association with charged leptons, with
mixing angles of order unity. LHC will be
sensitive to both mechanisms \cite{pil4}, but the most 
stringent limits are expected from the latter processes \cite{interplay,LHC}.
On the other hand, $e^+e^-$ colliders will be mainly sensitive to neutrino pair
production \cite{clic04,CLICpair}, which might be used to learn about heavy
neutrino properties as well \cite{axel}. We note that in the case of Majorana
neutrinos, $NN$ production may give an interesting lepton number violating
final state signal $NN \to \ell^\pm W^\mp \ell^{'\pm} W^\mp$ \cite{lnv,lnv2},
which has smaller backgrounds than the analogous lepton number conserving final
state $\ell^\pm W^\mp \ell^{'\mp} W^\pm$. In case that new interactions are
pushed to high energies, heavy neutral leptons can still be produced in 
pairs if they transform non-trivially under the SM gauge group \cite{nprod}.

We have discussed the prospects for TeV scale neutrino detection
in $\ell W \nu$ production at future $e^+ e^-$ colliders, taking as example
the CLIC proposal for a 3 TeV collider at CERN. This study complements the
analysis of the CLIC potential based on new gauge interactions
\cite{clic04,CLICpair}. We have examined the
dependence of the results on the heavy neutrino mass, and special emphasis has
been made on the importance of a complete study of the three possible channels 
$\ell = e,\mu,\tau$. As it has been argued, a non-negligible coupling to the
electron, $V_{eN} \sim 10^{-3} - 10^{-2}$ for $m_N = 1 - 2$ TeV, is necessary
to produce the heavy neutrino at detectable rates. The produced neutrinos can
then decay $N \to eW$ and, if they couple to the muon and tau, $N \to \mu W$
and $N \to \tau W$ as well (plus additional decays mediated by neutral
currents). Among these, the muon channel is by far the cleanest one due to its
smaller background. We have
explicitly shown that for $m_N = 1500$ GeV the sensitivity to $V_{eN}$ improves
from $V_{eN} \sim 0.008$ to $V_{eN} \sim 0.0035$ in the presence of a small
coupling $V_{\mu N} \sim 0.005$, but is hardly affected by $N$ mixing with
the tau lepton.

We have also studied what we could learn about heavy neutrinos if they were
discovered at CLIC. It has been shown that the angular distribution of the
produced neutrino relative to the beam axis is a powerful discriminant between
Majorana and Dirac neutrinos, giving a clear evidence of their nature even for
relatively small signals. Then, we have discussed how to extract heavy neutrino
charged current couplings from data, estimating with an example
the precision with which they could be eventually measured. We have proposed a
method to determine the chiral structure of the $\ell N W$ interactions, which
unfortunately is only useful for neutrino masses below the TeV scale.

Finally, we have performed a similar analysis for heavy neutrinos of few
hundreds of GeV at ILC, extending a recent analysis \cite{paper1}. In
particular,
we have explicitly investigated the flavour and mass dependence of the limits on
neutrino mixing with charged leptons. For a heavy neutrino with a mixing large
enough, we have shown how to establish its Dirac or Majorana nature and how to
determine the chirality of its charged current couplings.

The results obtained show that CLIC would outperform previous
machines in finding direct or indirect signals of heavy neutrinos. It would
extend direct searches up to masses around 2.5 TeV, and for masses around 1
TeV it would be sensitive to couplings $V_{eN} \simeq 4 \times 10^{-3}$. If no
heavy neutrino signal was found at CLIC, the bound
$V_{eN} \leq 2 - 6 \times 10^{-3}$ 
could be set for $m_N = 1 - 2$ TeV, improving the present limit
$V_{eN} \leq 0.073$ by more than one order of magnitude and matching a future
limit obtained at the Giga$Z$ option of ILC \cite{tdr}, for which a $10^{3}$
statistical gain would be expected to translate into bounds $\sim 30$ times
more stringent than those of Eqs.~(\ref{eps1}).
The direct limit on the product $V_{eN} V_{\mu N}^*$ obtained from the
non-observation of heavy neutrinos would be much more restrictive than the
present
indirect bound from LFV processes, and would remain competitive with future
improvements of the upper bounds on $\mathrm{Br}(\mu \to e \gamma)$ \cite{MEG}
and $\mu - e$ conversion in nuclei \cite{MECO}. For heavy neutrinos with masses
of few hundreds
of GeV, CLIC would probe mixing angles $V_{eN} \sim 10^{-3}$, one order of
magnitude better than what will be achieved at ILC.

\vspace{1cm}
\noindent
{\Large \bf Acknowledgements}

\vspace{0.4cm} \noindent
F.A. thanks J. Bernab\'eu, A. Bueno, J. Wudka and M. Zra{\/l}ek for discussions.
J.A.A.S. thanks F. R. Joaquim for useful comments.
This work has been supported in part
by MEC and FEDER Grant No. FPA2003-09298-C02-01,
by Junta de Andaluc{\'\i}a Group FQM 101,
by FCT through projects POCTI/FNU/44409/2002, CFTP-FCT UNIT 777 and
grant SFRH/BPD/12603/2003, and by the European Community's
Human Potential Programme under contract HPRN-CT-2000-00149 Physics at
Colliders.

\end{document}